\documentclass[fleqn,usenatbib]{mnras}
\usepackage{newtxtext,newtxmath}
\usepackage[T1]{fontenc}

\DeclareRobustCommand{\VAN}[3]{#2}
\let\VANthebibliography\thebibliography
\def\thebibliography{\DeclareRobustCommand{\VAN}[3]{##3}\VANthebibliography}

\usepackage{graphicx}	% Including figure files
\usepackage{amsmath}	% Advanced maths commands

\newcommand*  {\diff}       {\mathop{}\!\mathrm{d}}
\renewcommand*{\vec}[1]     {\boldsymbol{#1}}
\newcommand*  {\uvec}[1]	{\hat{\vec{#1}}}
\newcommand*  {\p}          {\partial}
\newcommand*  {\phm}        {\phantom{-}}
\newcommand*  {\pdiff}[2]   {\frac{\p{#1}}{\p{#2}}}

\newcommand*  {\sub}[2]		{{#1}_{\mathrm{#2}}}
\title[Galactic bar models]
{A family of potential-density pairs for galactic bars}

\author[Dehnen, Aly]{%
\href{http://orcid.org/0000-0001-8669-2316}{Walter~Dehnen\includegraphics[width=11pt]{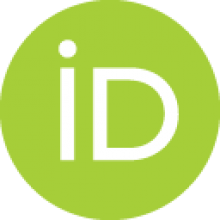}},$^{\!\!1,2}$ and
\href{http://orcid.org/0000-0002-1342-1694}{Hossam~Aly\includegraphics[width=11pt]{orcid-ID}}$^{1}$
\smallskip
\\
$^1$ Astronomisches Recheninstitut, Zentrum f{\"u}r Astronomie der Universit{\"a}t Heidelberg, M{\"o}nchhofstra\ss{}e 12-14, 69120, Heidelberg, Germany\\
$^2$ School for Physics and Astronomy, University of Leicester, University Road, LE1 7RH, UK
}

\date{Accepted XXX. Received YYY; in original form ZZZ}

\pubyear{2022}

\begin{document}

\defcitealias{MiyamotoNagai1975}{MN75}
\defcitealias{LongMurali1992}{LM92}
\allowdisplaybreaks
%%%%%%%%%%%%%%%%%%%%%%%%%%%%%%%%%%%%%%%%%%%%%%%%%%%%%%%
\label{firstpage}
\pagerange{\pageref{firstpage}--\pageref{lastpage}}
\maketitle

\begin{abstract}
We present a family of analytical potential-density pairs for barred discs, which can be combined to describe galactic bars in a realistic way, including boxy/peanut components. We illustrate this with two reasonable compound models. Computer code for the evaluation of potential, forces, density, and projected density is freely provided.
\end{abstract}


%%%%%%%%% SECTION %%%%%%%%%
\section{Introduction}
A large fraction of disc galaxies are barred \citep{EskridgeEtal2000, MenendezDelmestreEtAl2007, ShethEtAl2008, Erwin2018}, including our own Milky Way \citep[see the review by][and references therein]{BlandHawthornGerhard2016}. Bars are fascinating phenomena, which arise naturally from stellar discs via dynamic instability \citep[e.g.][]{BinneyTremaine2008}. They are important in re-distributing angular momentum between the stars in the disc, the surrounding dark-matter halo, and the cold gas \citep{Athanassoula2002, Athanassoula2003, WeinbergKatz2002, KormendyKennicutt2004, HolleyBockelmannEtal2005, DebattistaEtAl2006, Sellwood2014:review}, which is funnelled into the inner region, where it may drive star formation and an AGN. This re-distribution is believed to be the main driver for the secular evolution of barred disc galaxies. The study of these processes has become an important subject of research, in particular in the context of the Milky Way, where the ever richer data, especially owing to ESA's Gaia mission, provide us with a unique opportunity to study secular evolution in minute detail.

Theoretical studies of bar dynamics require some model for the gravitational potential of the bar. Investigations of stellar orbits have traditionally represented the bar component with a \cite{Ferrers1877} bar \citep[][and many subsequent studies]{Pfenniger1984}, which has unrealistic density profile and requires numerical computations. \cite{Dehnen2000:OLR} added a quadrupole term to an axisymmetric potential to study near-resonant orbits outside a bar. This approach is simple and fast, but only suitable for orbits that stay in the equatorial plane and outside the bar (though many subsequent studies have used this approach for inner-bar orbits).

Another, method is to compute the gravitational potential numerically either from an observationally motivated density model \citep[][and several subsequent studies]{PatsisAthanassoulaQuillen1997, HafnerEtal2000} or from an $N$-body model \citep*[e.g.][]{HarsoulaKalapotharakos2009,WangEtAl2020}. This option is often computationally expensive, in particular if the potential is computed using many multipole components to resolve vertically thin  structures, and renders it less suitable for studies involving high numbers of orbit integrations.

Yet another approach is to stretch an axisymmetric model along the $x$-direction by convolving it with a function $f(x)$ to obtain a triaxial shape. \citeauthor{LongMurali1992} (\citeyear{LongMurali1992}, hereafter \citetalias{LongMurali1992}) and \cite{WilliamsEvans2017} used a box function $f(x)$ to convolve, respectively, the disc model of \citeauthor{MiyamotoNagai1975} (\citeyear{MiyamotoNagai1975}, hereafter \citetalias{MiyamotoNagai1975}) and the axisymmetric logarithmic potential. The resulting triaxial models are fully analytical  and have near-constant density along the bar major axis, but are not very realistic. \cite*{McGoughEtAl2020} used an exponential in $|x|$ to convolve an oblate Gaussian, obtaining a model with near-exponential and Gaussian density profiles along and perpendicular to the bar, respectively. Unfortunately, this model requires one-dimensional numerical quadrature for potential and forces.

There are, however, still no bar models known that are simple enough for their gravitational potentials to be fully analytical and at the same time realistic enough to be useful in quantitative studies of bar orbits and secular evolution. In this work, we present novel analytical bar models, which extend that of \citetalias{LongMurali1992} in mainly two ways: first, we use more axisymmetric models with steeper radial and/or vertical profiles, and second, we employ linear functions $f(|x|)$ to convolve those models with. The resulting triaxial models are still fully analytic but more flexible and more realistic than that of \citetalias{LongMurali1992}, in particular when combining several of them.

This paper is organised as follows. After Section~\ref{sec:bar} presents a technique to render known and novel axisymmetric disc models barred, the properties of these barred models are explored in Section~\ref{sec:bar:props}. Section~\ref{sec:compound} presents two illustrative multi-component models and Section~\ref{sec:conclude} summarises this work.

%%%%%%%%%%%%%%%%%
\begin{figure*}
    \begin{center}
	    \includegraphics[width=86mm]{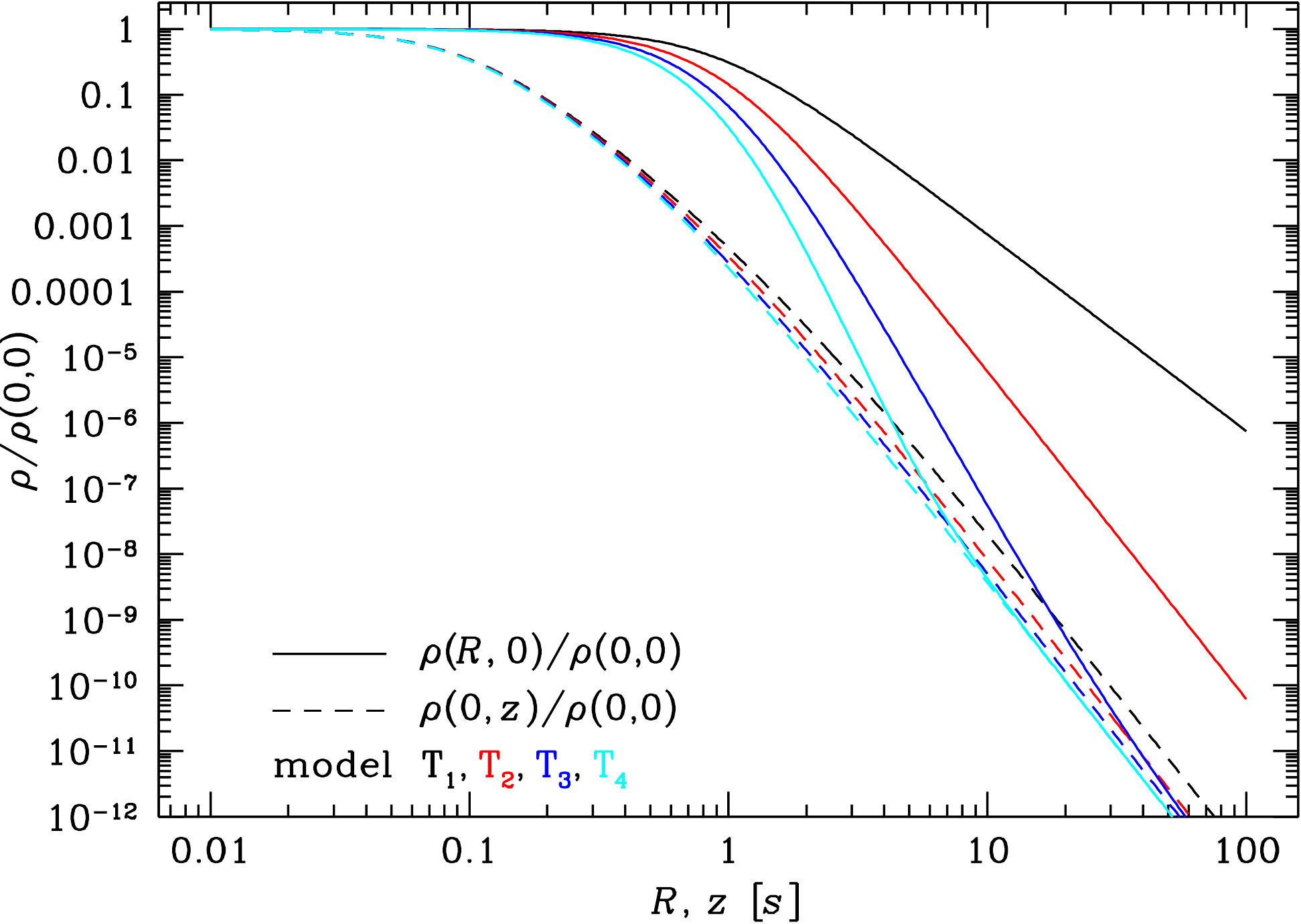}\hfill
	    \includegraphics[width=86mm]{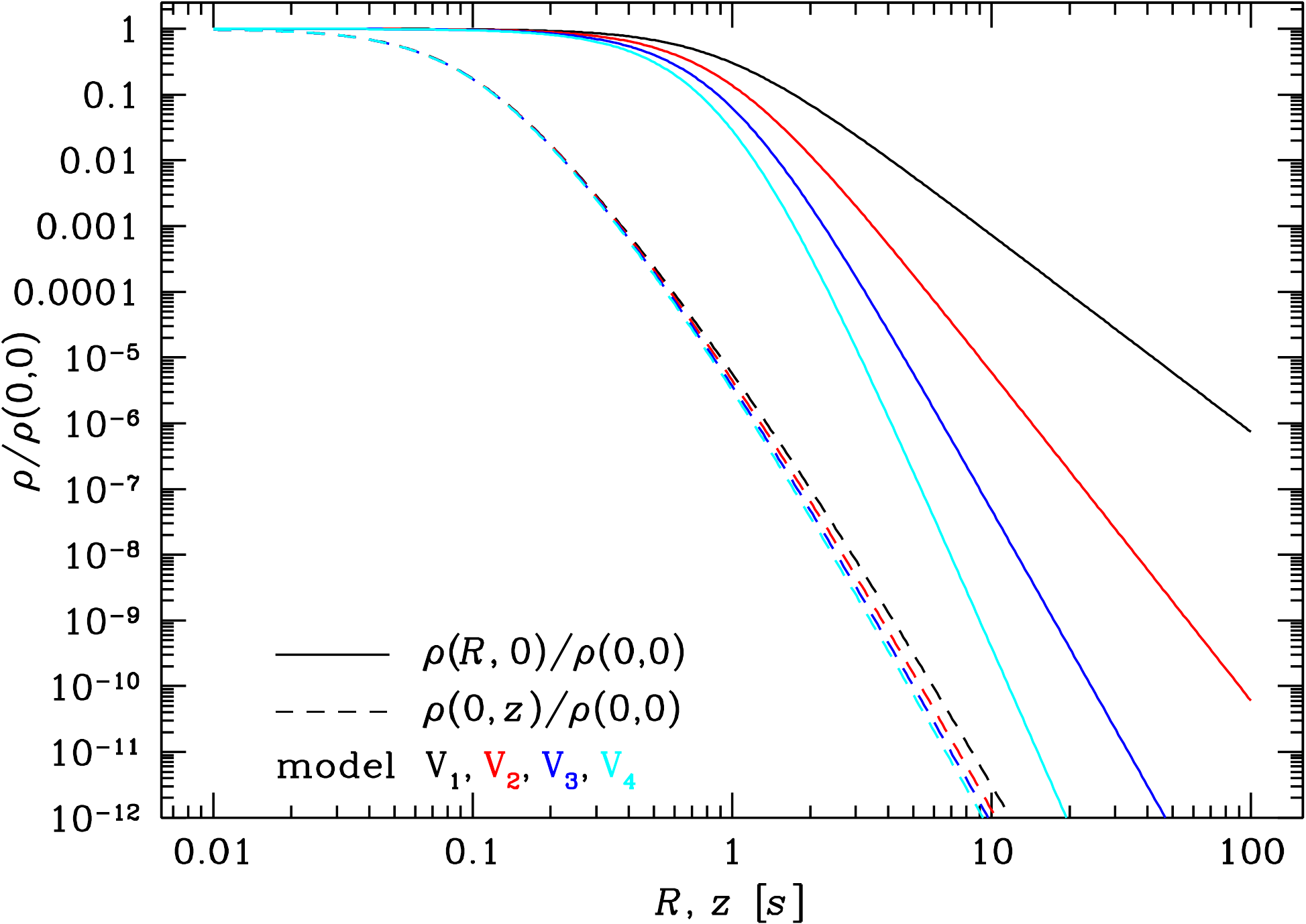}
    \end{center}
    \vspace*{-3mm}
	\caption{\label{fig:rho:TV}
	\textbf{Left:} radial density profile along the major (solid) and minor (dashed) axes for the axisymmetric models T$_k$ (with density given in equations~\ref{eqs:rho:Tk}) for $q=0.1$. Model T$_1$ is the standard Miyamoto-Nagai disc. \textbf{Right:} the same for the models V$_{\!k}$ (with density given in equations~\ref{eqs:rho:Vk}), which are obtained from the former via differentiation w.r.t.\ parameter $b$ and have almost identical major-axis profiles, but a significantly steeper vertical profile. All these models can be made barred by convolution with the needle density, which is accomplished by means of the replacements~\eqref{eq:bar:Phi,rho}.}
\end{figure*}
%%%%%%%%%%%%%%%%%
%%%%%%%%% SECTION %%%%%%%%%%%%%%%%%%%%%%%%%%%%%%%%%%%%%
\section{Barred disc models}
\label{sec:bar}
Our barred models are based on a family of axially symmetric models, the most basic of which is the well-known model of \citetalias{MiyamotoNagai1975}, which has potential and density
\begin{subequations}
    \label{eqs:T1}
\begin{align}
    \label{eq:pot:T1}
    \Phi(\vec{r}) & = -\frac{GM}{\varrho},
    \\
    \label{eq:rho:T1}
    \rho(\vec{r}) &= \frac{Mb^2}{4\upi\zeta^3} \left[\frac{a}{\varrho^3}+\frac{3 Z^2\zeta}{\varrho^5}\right],
\end{align}
\end{subequations}
where $\varrho=|\vec{\varrho}|$ with
\begin{align}
    \label{eq:varrho}
    \vec{\varrho} \equiv \{x,y,Z\} ,\quad
	Z \equiv \zeta + a,\quad
	\zeta\equiv\sqrt{z^2+b^2}.
\end{align}
Here, $b$ represents a scale height and $a$ a scale length. While these parameters are the most natural in terms of equations~\eqref{eqs:T1} and have been widely used, physically more meaningful parameters are arguably the scale radius $s=a+b$ and the dimensionless flattening or axis ratio
\begin{align}
    q=\frac{b}{a+b}\in[0,1],
\end{align}
when $a=s(1-q)$ and $b=sq$.

%%%%%%%%% SUB-SECTION %%%%%%%%%
\subsection{Making bars from axisymmetric discs}
\label{sec:bar:eqs}
%%%%%%%%%%%%%%%%%%%%%%%%%%%%%%%
Like \citetalias{LongMurali1992}, we convolve the \citetalias{MiyamotoNagai1975} model with an infinitely thin needle along the $x$-axis with unit mass and 3D density $\sub{\rho}{needle}(\vec{r})=f(x)\delta(y)\delta(z)$, where
\begin{align}
	\label{eq:f(r)}
    f(x) = \frac1{2L} \times \begin{cases}
        1 + \gamma(1-2 |x|/L) & \text{for $|x| < L$} \\
		0 & \text{otherwise}.
    \end{cases} 
\end{align}
Here, $L$ is the half-length (`radius') of the needle, while the parameter $\gamma\propto-\diff\sub{\rho}{needle}/\diff |x|$ extends the approach of \citetalias{LongMurali1992} to needles with density linear in $|x|$. For $-1\le\gamma\le1$, $f(x)$ remains non-negative everywhere and hence, by implication, also the density of the barred model constructed with it. We define the convolutions
\begin{align}
	\label{eq:In}
	\mathcal{I}_n(\vec{r};L,\gamma) \equiv 
	\sub{\rho}{needle}(\vec{r}) \ast r^{-n} \quad \text{with integer $n>0$},
\end{align}
which can be expressed in closed form as detailed in Appendix~\ref{app:In} (in fact, the convolution integrals are analytical for any piece-wise polynomial $f(x)$). We then have for the convolved \citetalias{MiyamotoNagai1975} model
\begin{subequations}
	\label{eqs:bar:needle:B}
\begin{align}
	\label{eq:bar:needle:B:Phi}
	\Phi(\vec{r}) &= -GM\,\mathcal{I}_{1}(\vec{\varrho};L,\gamma),
	\\
	\label{eq:needle:bar:B:rho}
	\rho(\vec{r}) &= 
	\frac{M b^2}{4\upi\zeta^3}\left[a\,\mathcal{I}_{3}(\vec{\varrho};L,\gamma) + 3Z^2\zeta\,\mathcal{I}_{5}(\vec{\varrho};L,\gamma)\right].
\end{align}
\end{subequations}
For the simplest case of a constant density along the needle ($\gamma=0$), equation~\eqref{eq:bar:needle:B:Phi} becomes
\begin{align}
	\Phi(\vec{r}) &= -\frac{GM}{2L}\ln\frac{x+L+\sqrt{\smash[b]{(x+L)^2+y^2+Z^2}}}{x-L+\sqrt{\smash[b]{(x-L)^2+y^2+Z^2}}},
\end{align}
equivalent to equation~(8b) of \citetalias{LongMurali1992}. Of course, any axisymmetric model can be transformed to become barred by convolution with the needle. If the model depends on $x$ only through integer powers $1/\varrho^n$, this convolution is accomplished by replacing
\begin{align}
	\label{eq:bar:Phi,rho}
	1/{\varrho^{n}} &\to \mathcal{I}_n(\vec{\varrho};L,\gamma)
\end{align}
in expressions for potential and density. We now present more axisymmetric models, to which this technique can be applied.

%%%%%%%%% SUB-SECTION %%%%%%%%%
\subsection{More axisymmetric disc models to be made barred}
\label{sec:more:models}
%%%%%%%%%%%%%%%%%%%%%%%%%%%%%%%
\citetalias{MiyamotoNagai1975} constructed their model by modifying the razor-thin \cite{Kuzmin1956} disc, which has density $\rho=\delta(z)\Sigma(R)$ with surface density
\begin{align}
    \Sigma(R) = \frac1{2\upi}\frac{Ma}{(R^2+a^2)^{3/2}}
\end{align}
and corresponds to the limit $b,\,q\to0$ of \citetalias{MiyamotoNagai1975}'s model. As \cite{Toomre1963} pointed out, models with the steeper surface density profiles
\begin{align}
	\label{eq:Toomre}
	\Sigma_{\mathrm{T}_k}(R) = 
	\frac{2k-1}{2\upi} \frac{Ma^{2k-1}}{(R^2+a^2)^{k+1/2}}
	\qquad\text{with $k\ge 1$}
\end{align}
can be obtained via differentiation with respect to the parameter $a$. The subscript `T$_k$' here stands for `Toomre's model $k$', such that \citeauthor{Kuzmin1956}'s disc is model T$_1$. These razor-thin models are the limits $b,\,q\to0$ of associated Toomre-Miyamoto-Nagai models with finite thickness, whose gravitational potentials and densities are derived in Appendix~\ref{app:Tk}.

%%%%%%%%%%%%%%%%%
\begin{figure*}
    \begin{center}
	    \includegraphics[height=56mm]{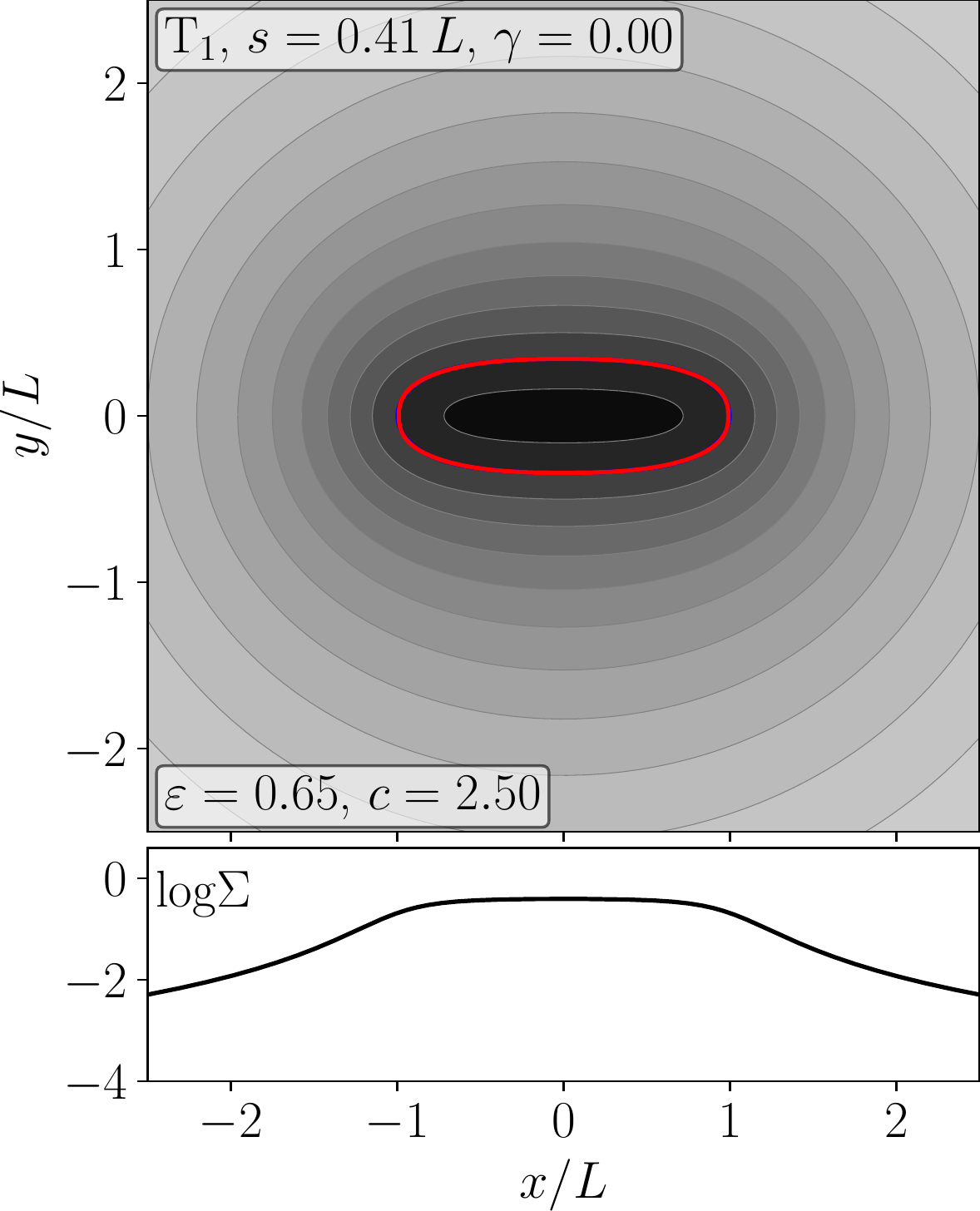}\hfil
	    \includegraphics[height=56mm]{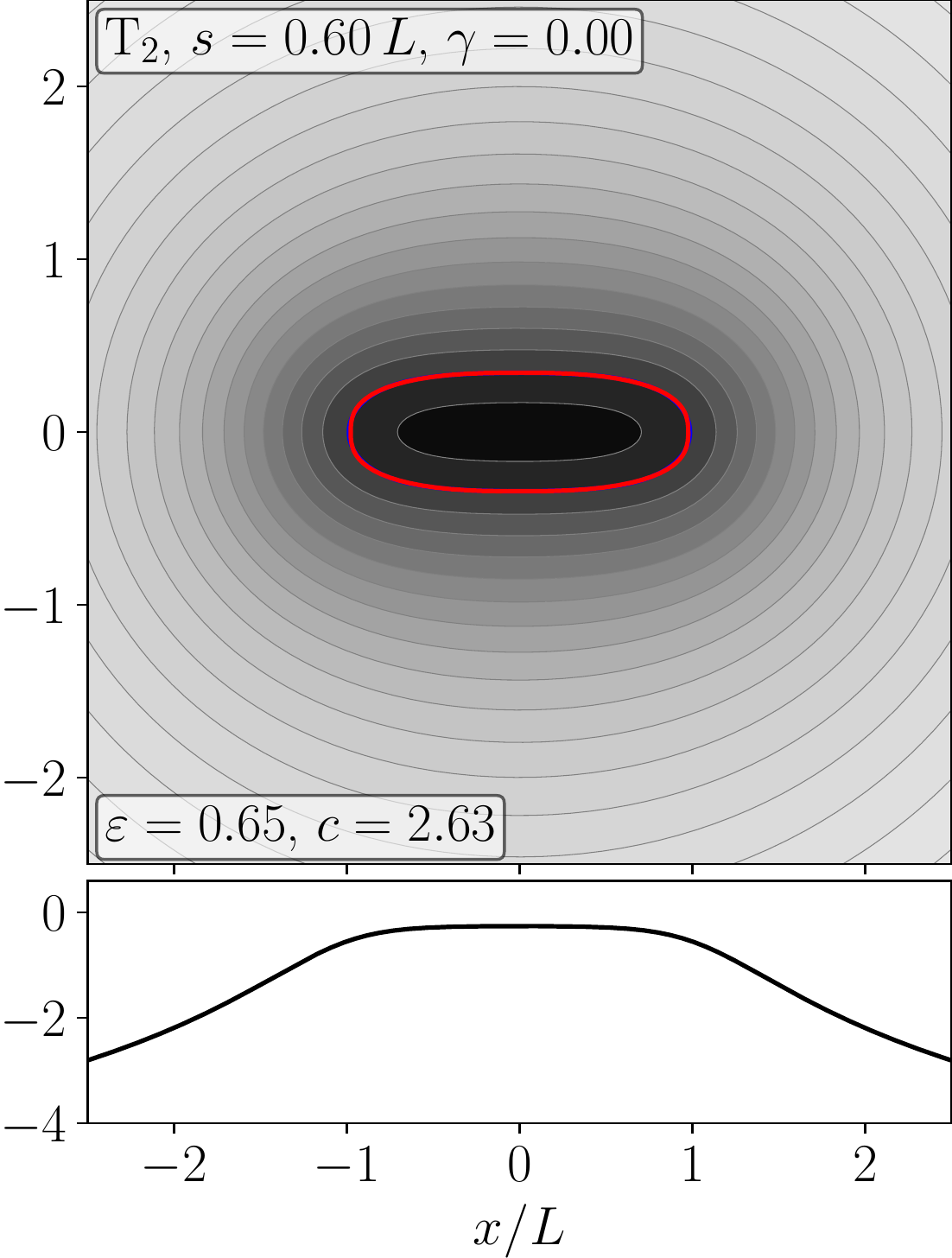}\hfil
 	    \includegraphics[height=56mm]{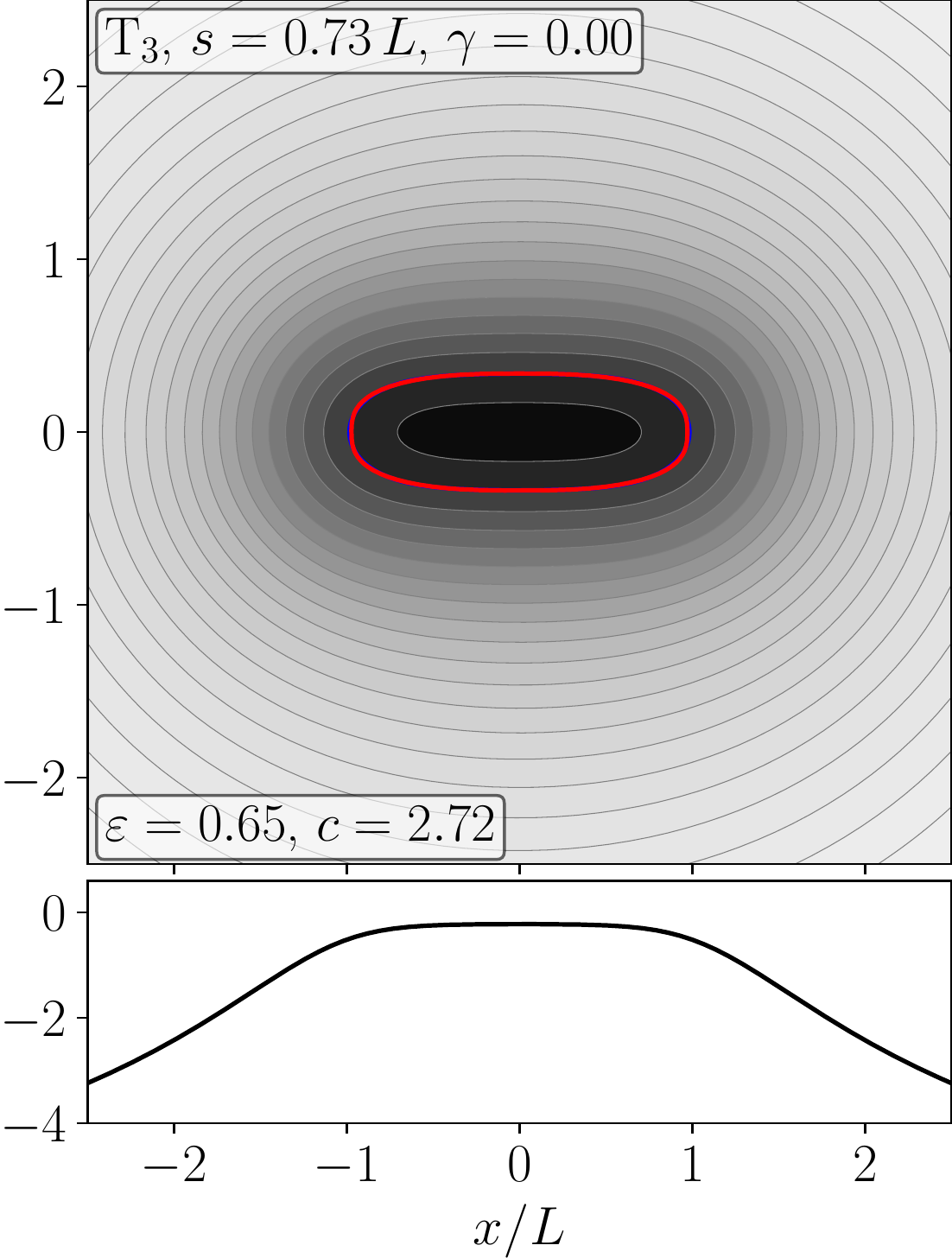}\hfil
 	    \includegraphics[height=56mm]{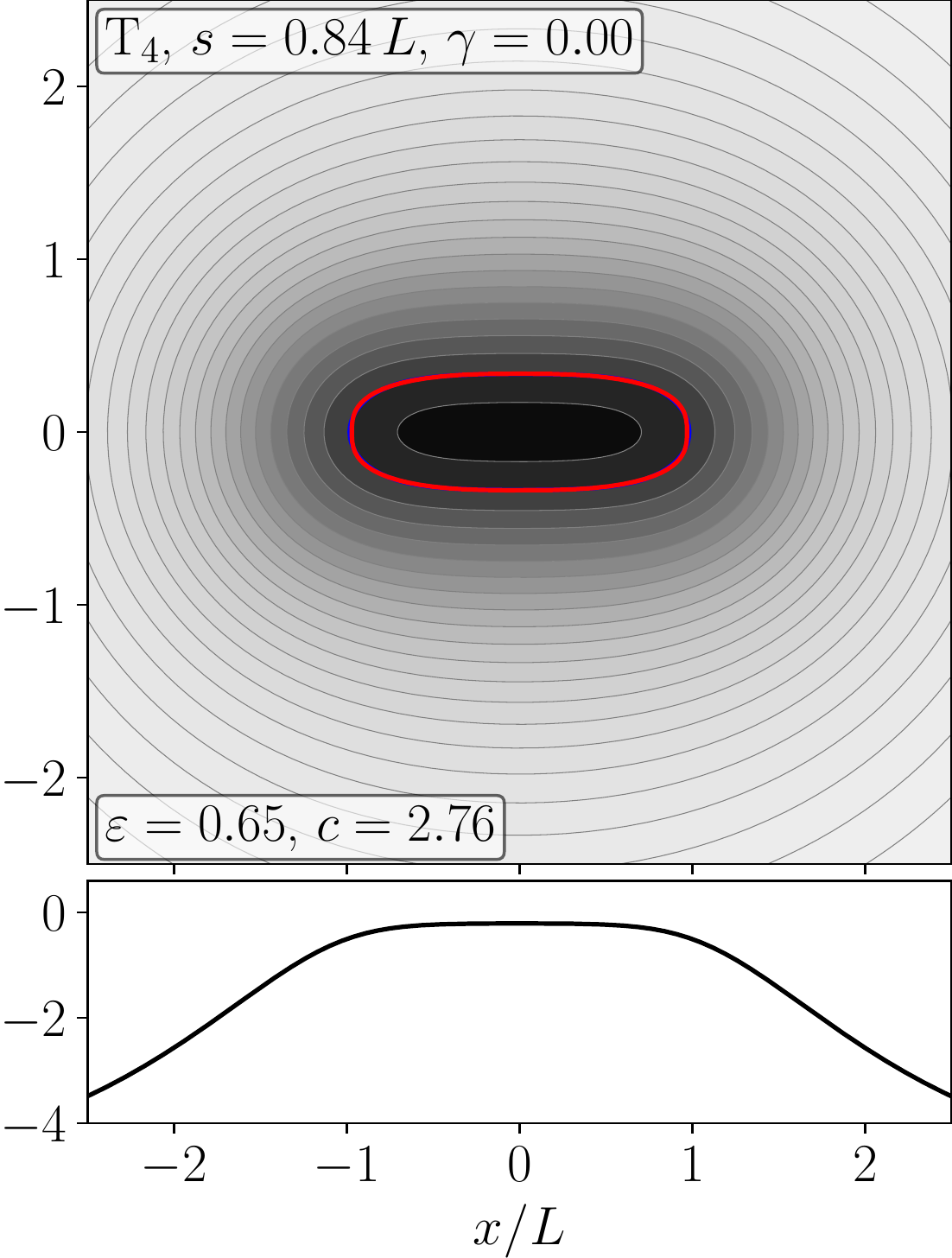}
    \end{center}
    \vspace*{-3mm}
	\caption{\label{fig:Sigma:bar:K}
	Surface density $\Sigma(x,y)$ (upper plots, normalised by its central value using contours equidistant in $\log\Sigma$) and along the major axis (lower plots) of the barred models T$_{\text{1-4}}$ for $q=0.1$ and $\gamma=0$. The ratios $s/L$ between scale radius and bar radius have been adjusted such that the bars have similar axis ratios. For the contour with semi-major axis equal to $L$, we perform a generalised ellipse fit (red, see equation~\ref{eq:general:ellipse}) and report its ellipticity $\varepsilon=1-a_y/a_x$ and boxiness $c$ in the bottom left of each plot. The model in the leftmost panel is the same as the `triaxial bar' of \protect\citetalias{LongMurali1992}.
	}
\end{figure*}
%%%%%%%%%%%%%%%%%

The left panel of Fig.~\ref{fig:rho:TV} plots the density profiles along the major and minor axes, respectively, of the models T$_{\text{1-4}}$ with axis ratio $q=0.1$. At large radii, model T$_1$ decays as $R^{-3}$ and $|z|^{-5}$, respectively. For models T$_{k>1}$, the behaviour on the minor axis remains very similar, but on the major axis they decay increasingly faster, though asymptotically all reach $R^{-5}$. In the spherical limit ($a=0$ or $q=1$) all these models  become the \cite{Plummer1911} sphere $\rho\propto(r^2+b^2)^{-5/2}$.

Novel models with steeper vertical and radial density profiles can be obtained by differentiating the Toomre-Miyamoto-Nagai models T$_k$ with respect to the parameter $b$. Applying this once gives new models which we denote `V$_{\!k}$' and whose gravitational potential and densities are derived in Appendix~\ref{app:Vk}.

The density profiles along major and minor axes of models V$_{\!\text{1-4}}$ with $q=0.1$ are shown in the right panel of Fig.~\ref{fig:rho:TV}. They have vertical asymptote $\rho\sim |z|^{-7}$ at $|z|\gg b$, while their major-axis behaviour remains very similar to the associated T$_k$ model (though models V$_{k>2}$ ultimately approach $\rho\sim R^{-7}$ at large $R$). In the spherical limit ($a=0$ or $q=1$), these models approach the spherical model with $\rho\propto(r^2+b^2)^{-7/2}$, independently of $k$, while in the razor-thin limit ($b\to0$ or $q\to0$), they revert to the associated T$_k$ model. 

Models with yet steeper vertical profile, asymptoting to $|z|^{-9}$, can be obtained by differentiating once more with respect to $b$ (we refrain from this exercise here). All these models have finite vertical curvature $\p^2\rho/\p z^2$ everywhere. This is contrast to the exponential vertical density profile, $\rho\propto\exp(-|z|/h)$, which has infinite vertical curvature at $z=0$. In other words, unlike an exponential, the models are smooth and not spiky towards the mid-plane.

The gravitational potential $\Phi$ and density $\rho$ of all these axisymmetric models depend on $x$ only through integer powers $1/\varrho^n$ (see Appendix~\ref{app:axi}), such that barred versions can be obtained from the  recipe~\eqref{eq:bar:Phi,rho}. These barred models inherit the asymptotic behaviour at large $|z|$ and $R$ as well as the smooth behaviour at $z=0$ from their axisymmetric parent models. In the razor-thin limit, the barred models have surface density obtained by replacing
\begin{align}
    \label{eq:bar:Sigma}
	\frac1{(R^2+a^2)^{k+1/2}} &\to \mathcal{I}_{2k+1}\big(\{x,y,a\};L,\gamma\big)
\end{align}
in equation~\eqref{eq:Toomre}.

%%%%%%%%% SUB-SECTION %%%%%%%%%
\subsection{Projected density}
\label{sec:projection}
%%%%%%%%%%%%%%%%%%%%%%%%%%%%%%%
For comparison with observed galaxies, the density (of the finite-thickness models) must be projected onto the plane of the sky. For arbitrary lines of sight, the projection integral cannot be expressed in closed form and requires numerical treatment for the axisymmetric as well as the barred models. For the most important vertical (face-on) projection, a special numerical treatment allowing Gau\ss-Legendre integration and hence enabling efficient simultaneous computation of the projected density $\Sigma$ at many projected positions is detailed in Appendix~\ref{app:proj:z}.

For the end-on projection of the barred models, the surface density is, of course, the same as for the associated axisymmetric model, which can be expressed in closed form via
\begin{align}
	\label{eq:proj:x}
	\int_{-\infty}^\infty \frac1{\varrho^n}\,\diff x &= \frac{B(\tfrac12,\tfrac{n-1}2)}{(y^2+Z^2)^{(n-1)/2}},
\end{align}
where $B(x,y)$ denotes the beta function. The side-on projection is obtained in closed form via
\begin{align}
	\label{eq:proj:y}
	\int_{-\infty}^\infty \mathcal{I}_n(\vec{\varrho};L,\gamma) \diff y = B(\tfrac12,\tfrac{n-1}2)\;
		\mathcal{I}_{n-1}\big(\{x,0,Z\};L,\gamma\big).
\end{align}
Finally, the projection along an arbitrary horizontal line of sight, which has angle $\varphi\neq0$ with respect to the $x$-axis can be derived in the same way as for the side-on projection (which corresponds to $\varphi=\upi/2$). The convolution is just reduced to a shorter needle. This is achieved by replacing in equation~\eqref{eq:proj:y} $L\to L|\sin\varphi|$ as well as $x\to X\equiv x\sin\varphi-y\cos\varphi$ and $y\to Y\equiv y\sin\varphi+x\cos\varphi$, the horizontal coordinates perpendicular to and along the line of sight, respectively. The end-on projection~\eqref{eq:proj:x} is the limit of $\varphi\to0$.

%%%%%%%%%%%%%%%%%
\begin{figure*}
    \begin{center}
	    \includegraphics[height=56mm]{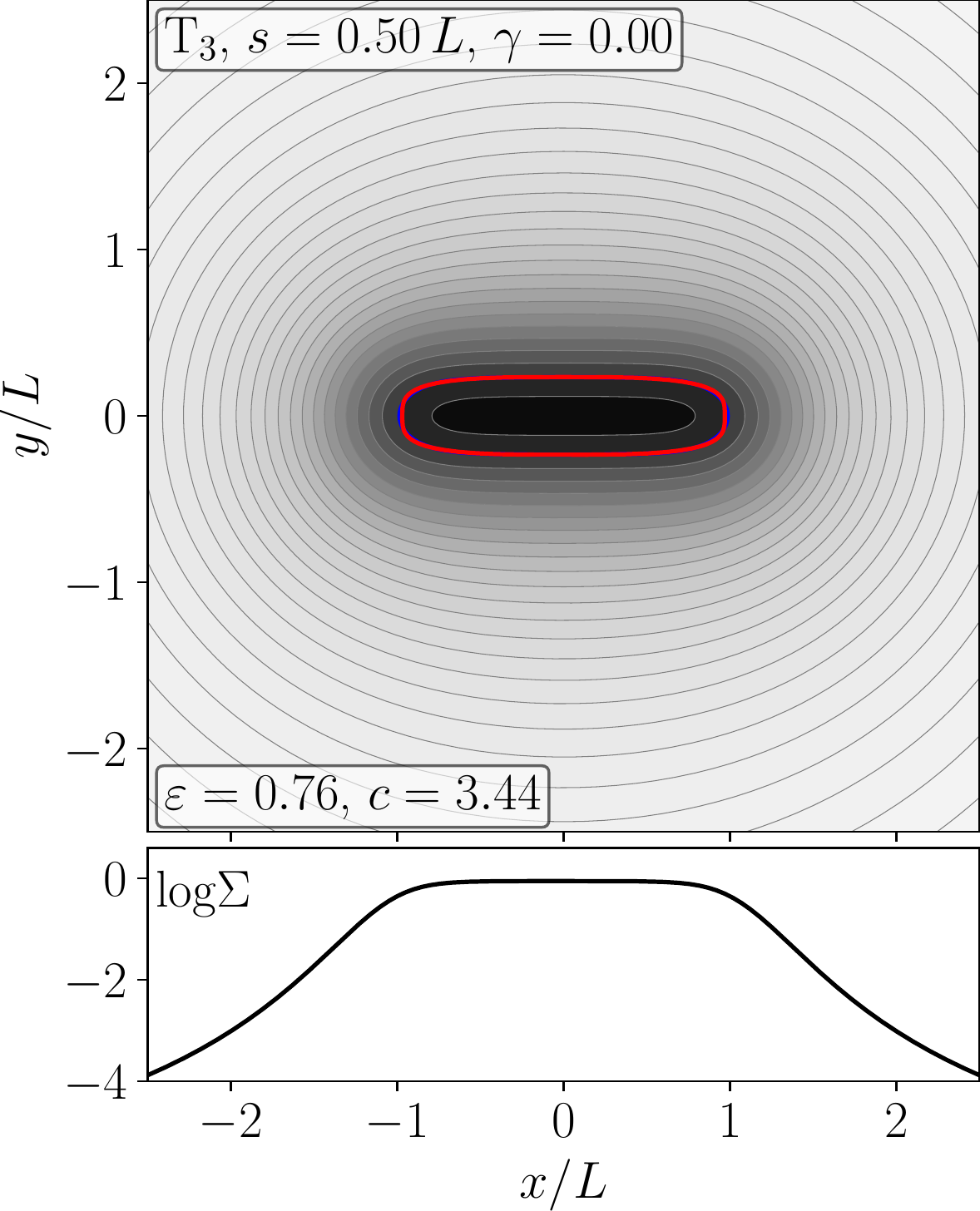}\hfil
	    \includegraphics[height=56mm]{figs/figSigBar_t3_073_p00_N.pdf}\hfil
 	    \includegraphics[height=56mm]{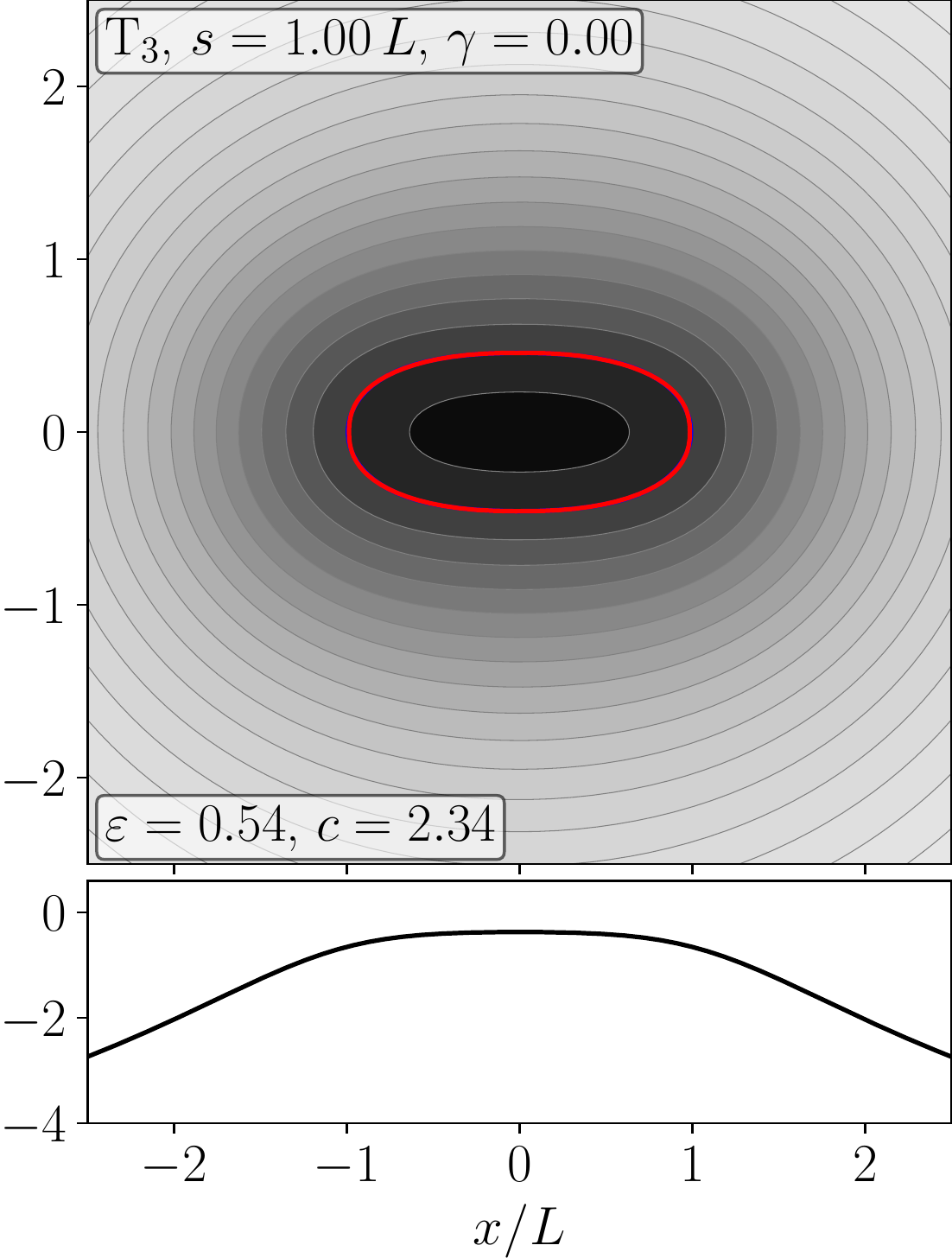}\hfil
 	    \includegraphics[height=56mm]{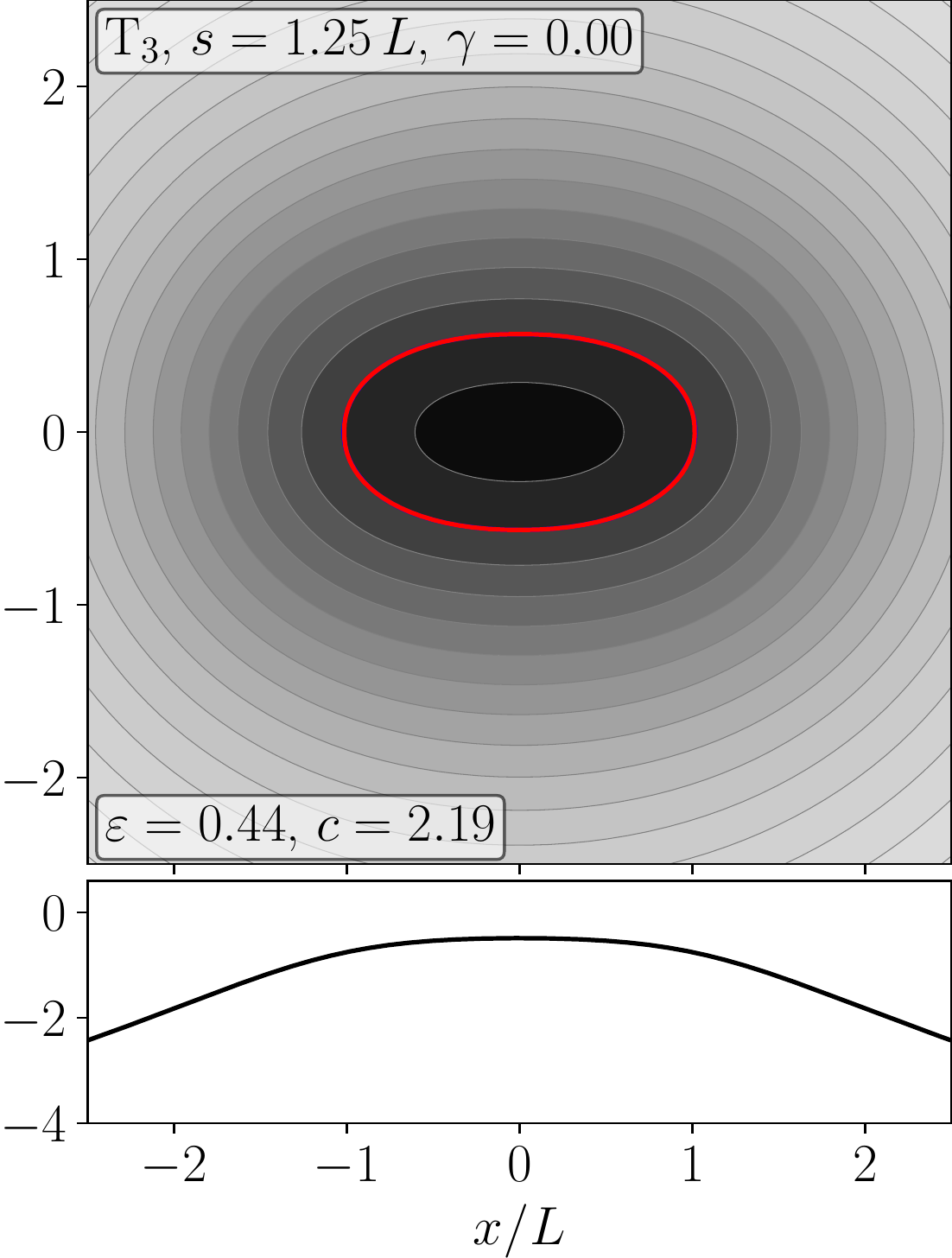}
    \end{center}
    \vspace*{-3mm}
	\caption{\label{fig:Sigma:bar:S}
	Like Fig.~\ref{fig:Sigma:bar:K} but for different $s/L$ (as indicated) for model T$_3$ with $q=0.1$ and $\gamma=0$. The second panel is identical to the third in Fig.~\ref{fig:Sigma:bar:K}.
	}
\end{figure*}
%%%%%%%%%%%%%%%%%
%%%%%%%%%%%%%%%%%
\begin{figure*}
    \begin{center}
        \includegraphics[height=56mm]{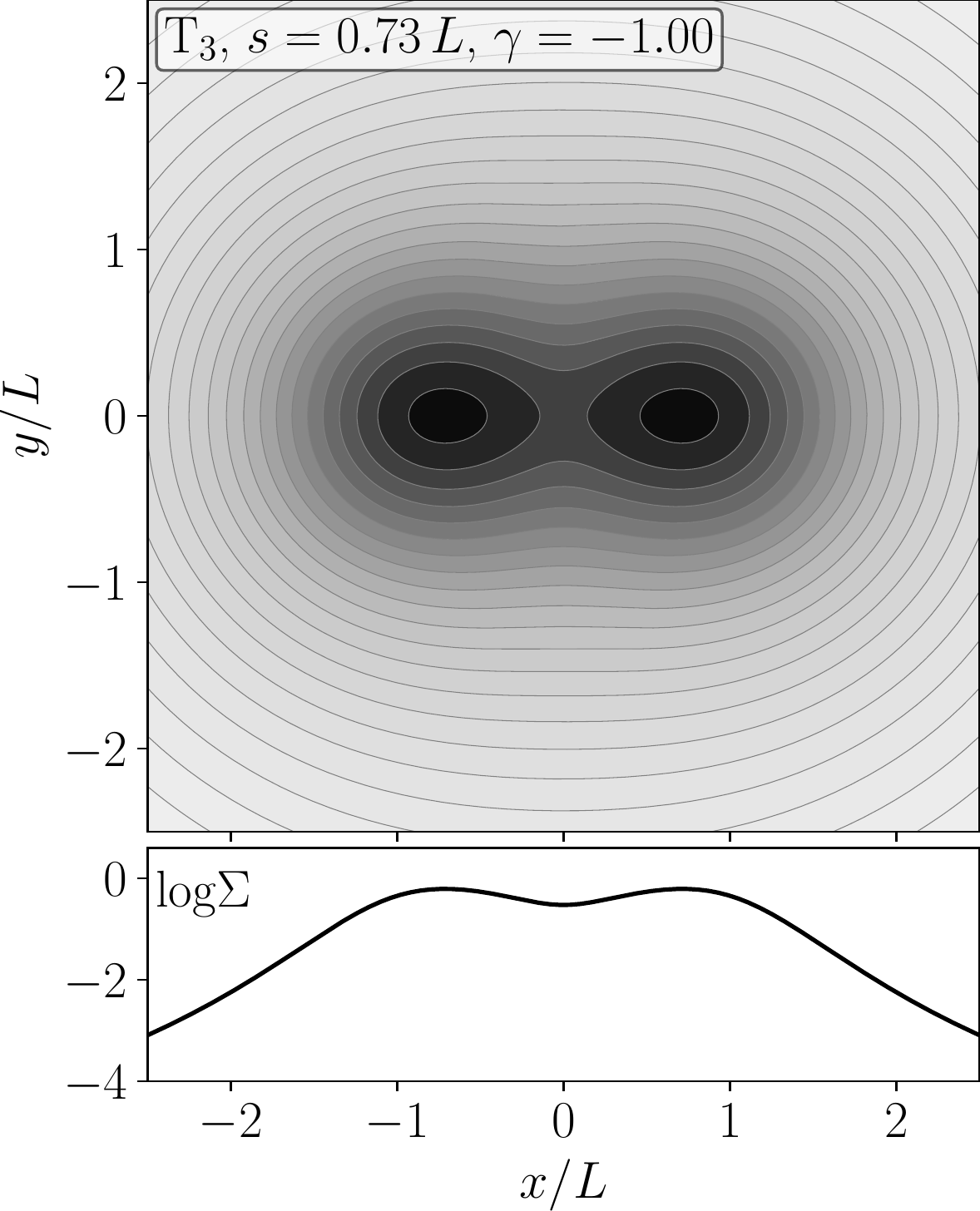}\hfil
	    \includegraphics[height=56mm]{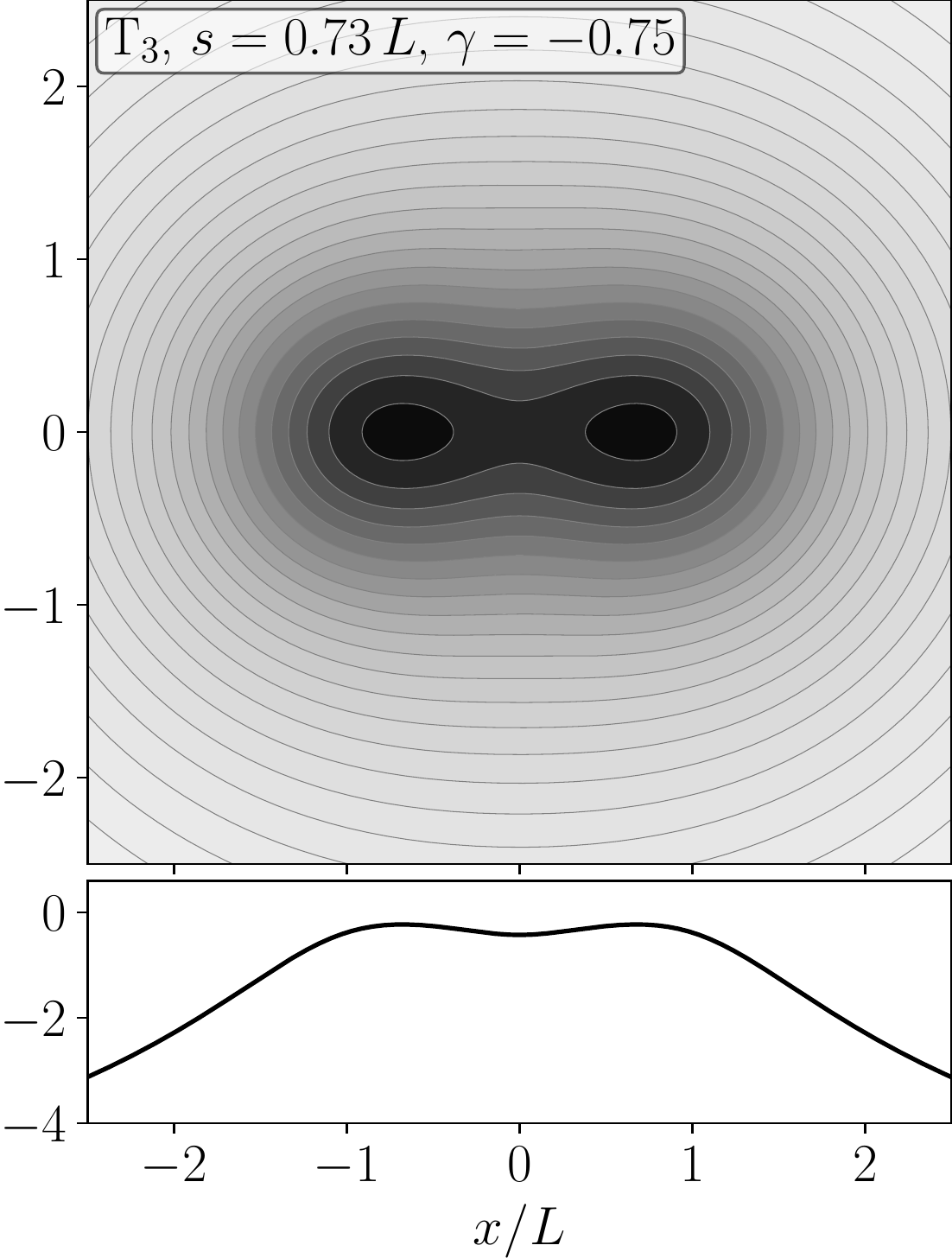}\hfil
 	    \includegraphics[height=56mm]{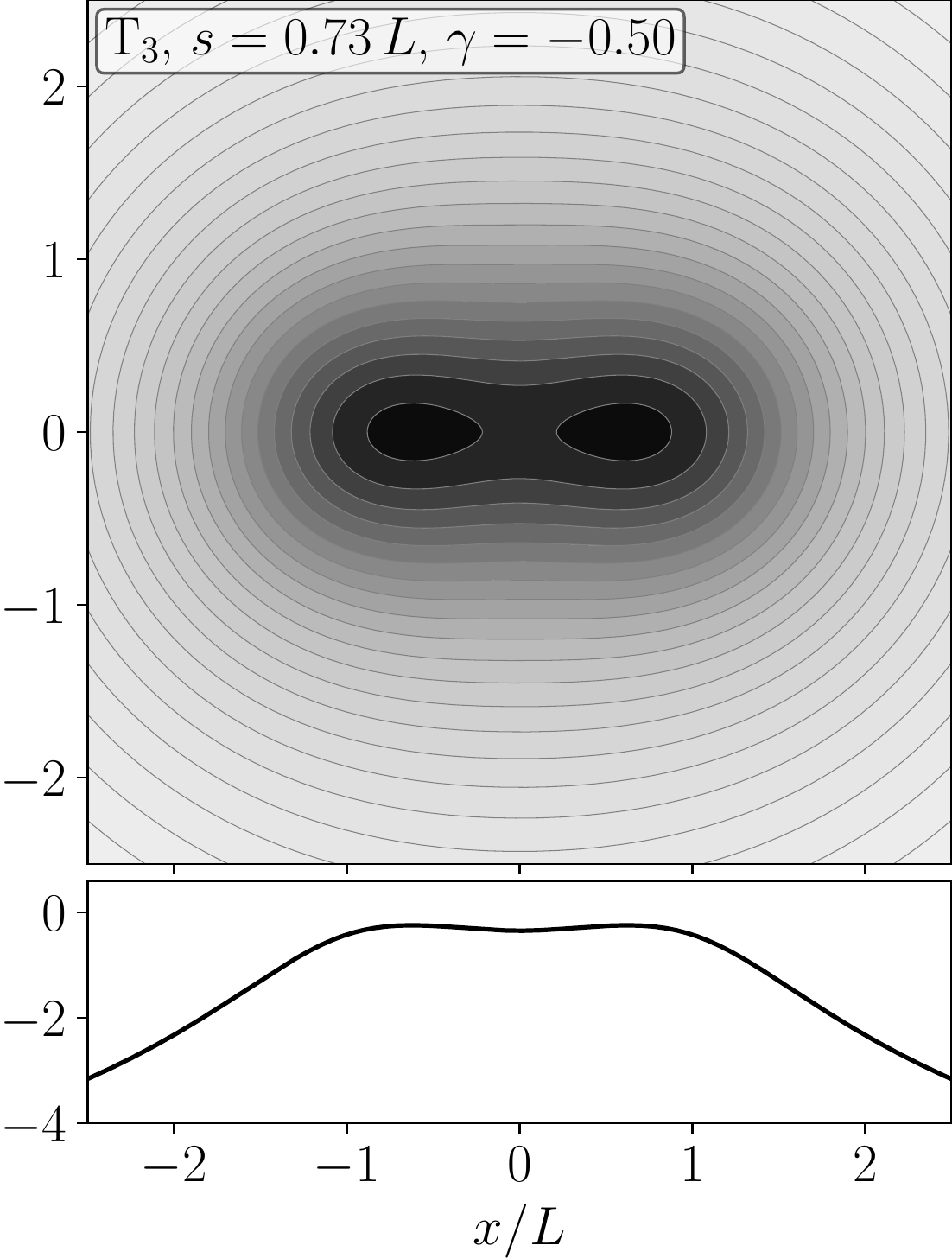}\hfil
 	    \includegraphics[height=56mm]{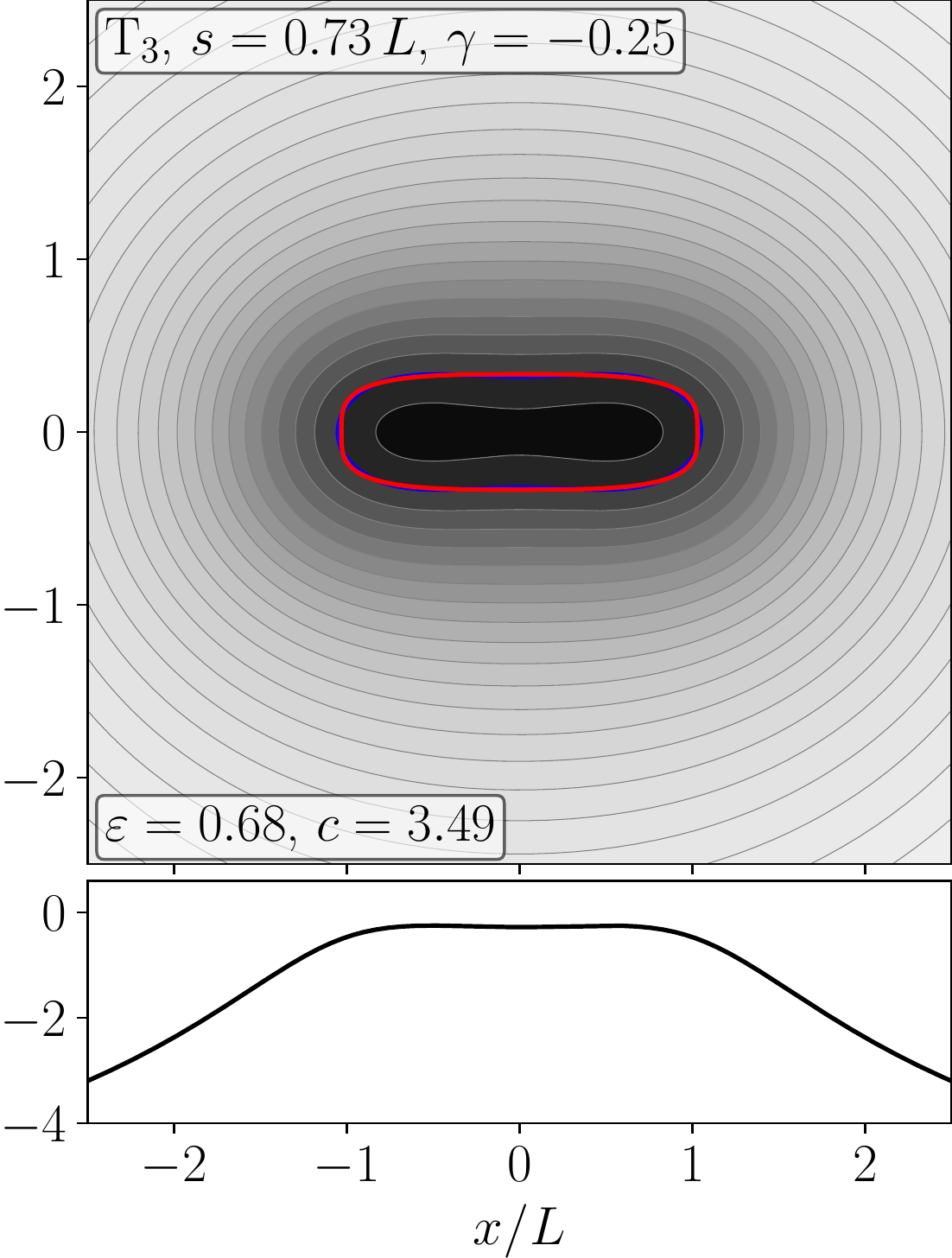}

	    \includegraphics[height=56mm]{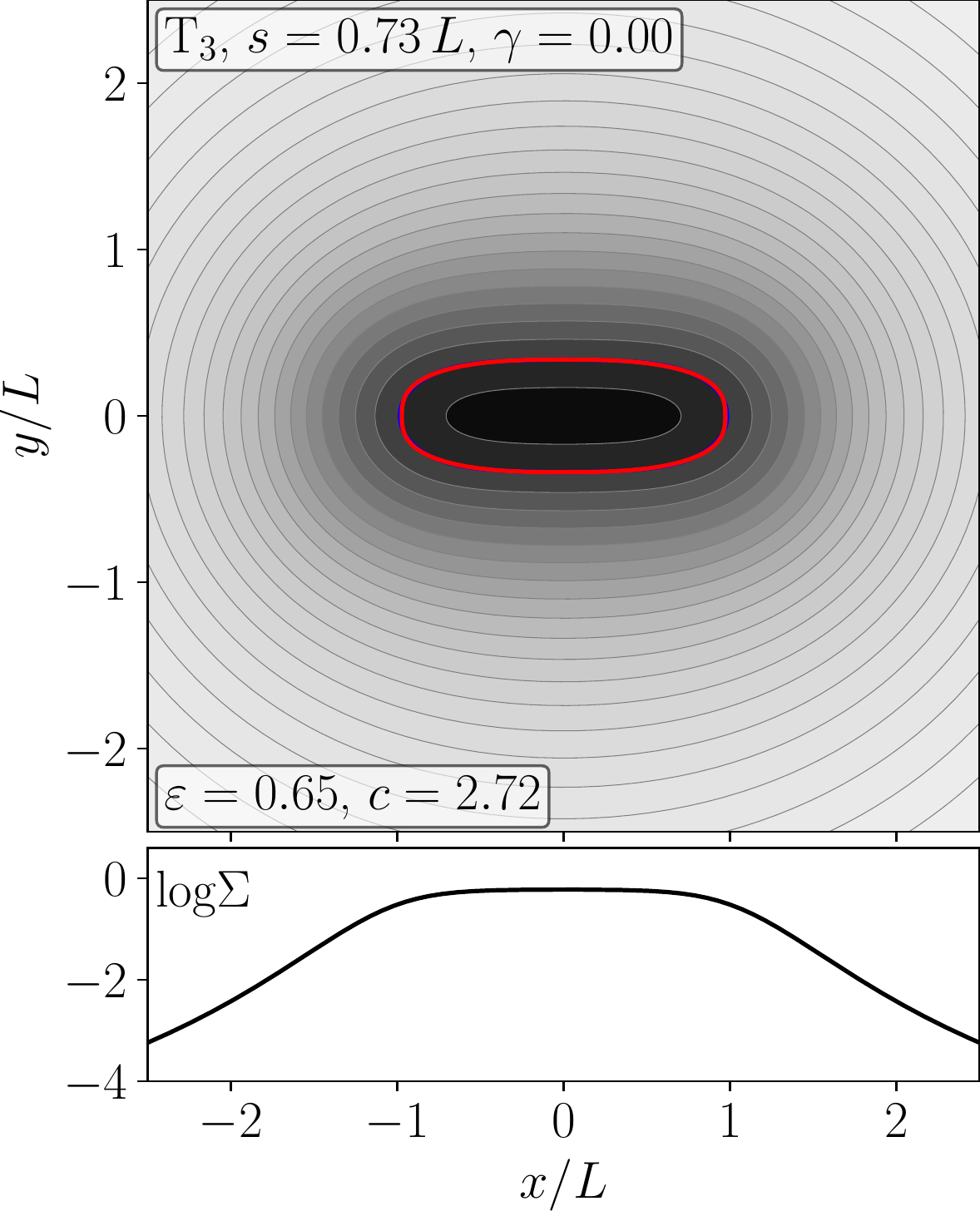}\hfil
	    \includegraphics[height=56mm]{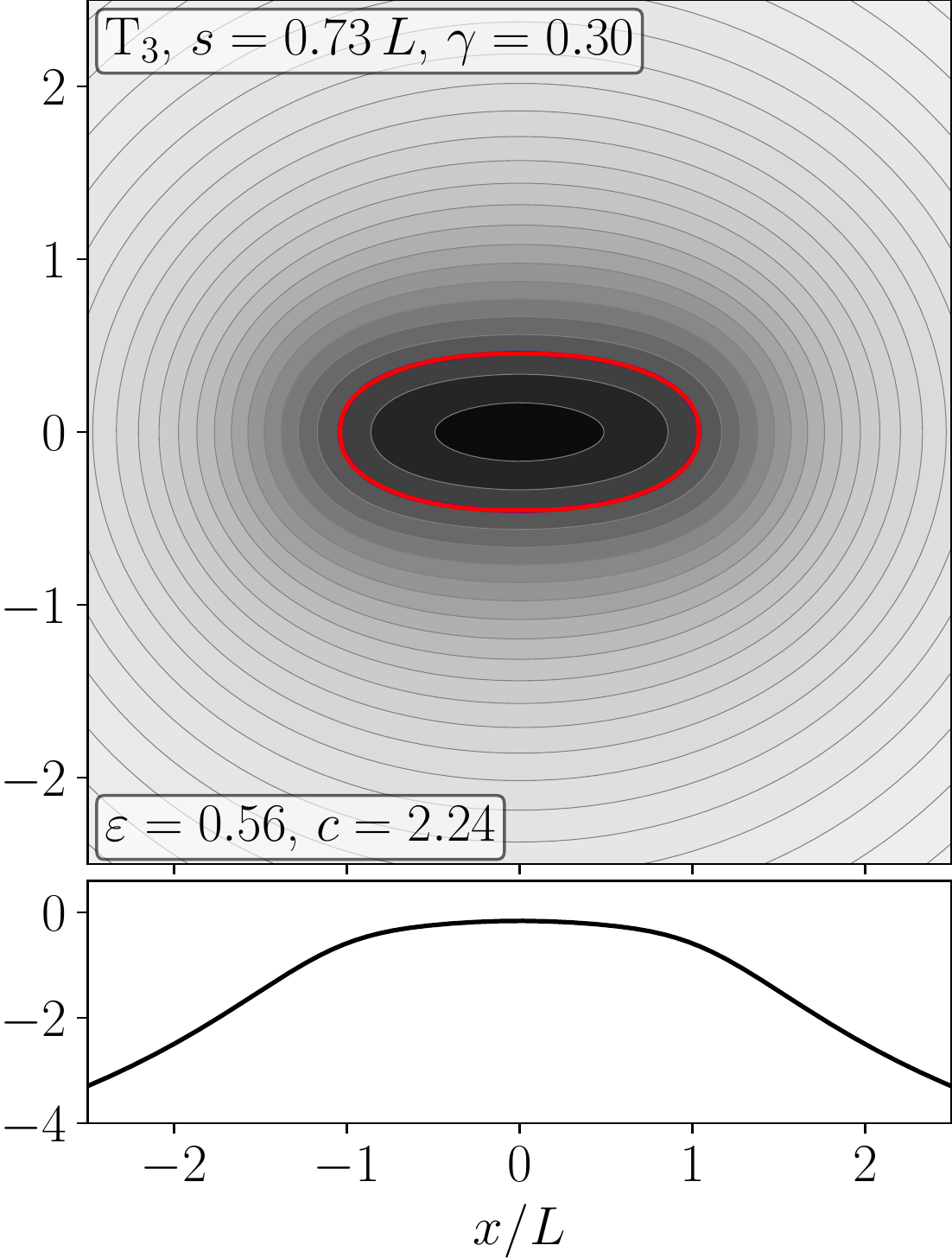}\hfil
 	    \includegraphics[height=56mm]{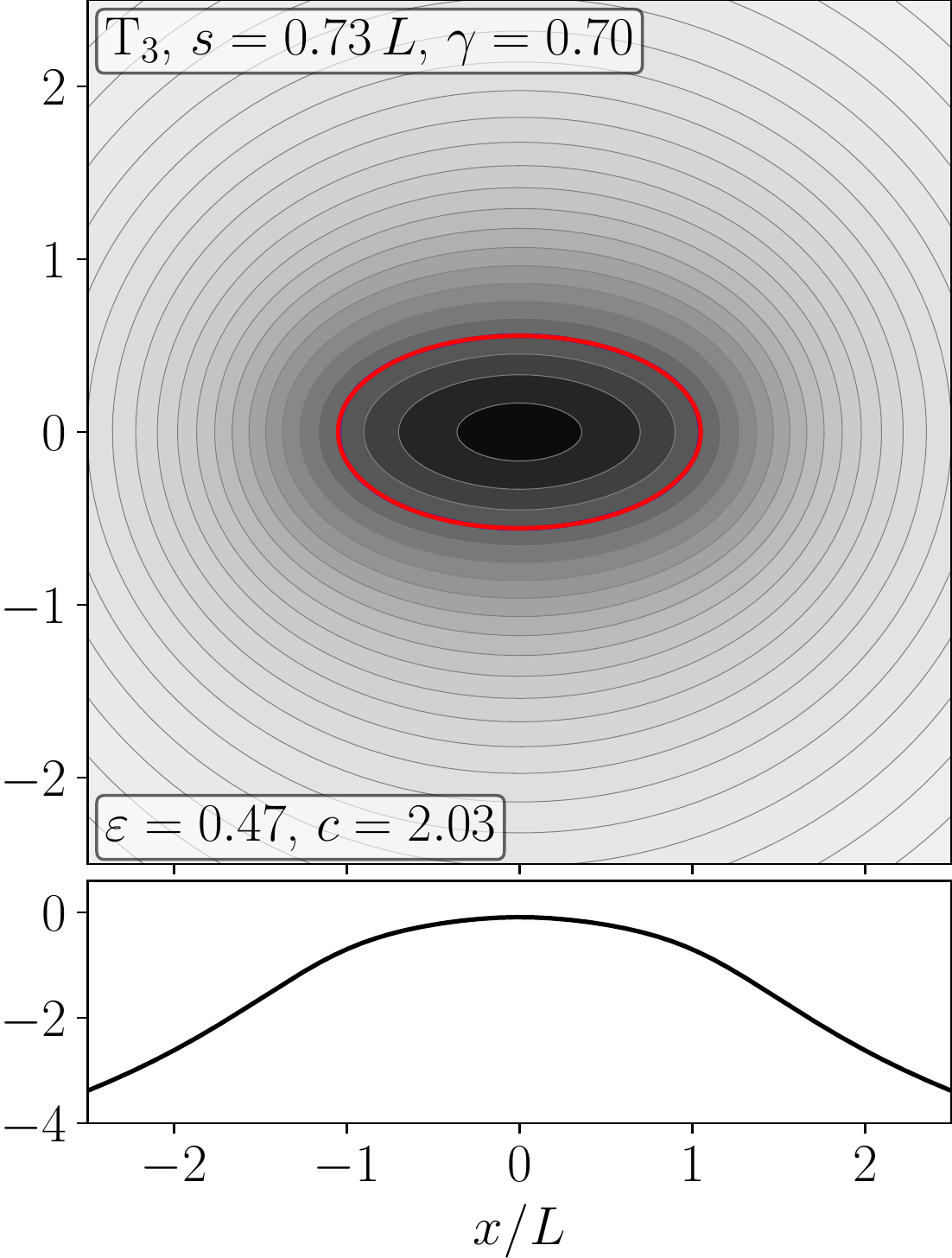}\hfil
 	    \includegraphics[height=56mm]{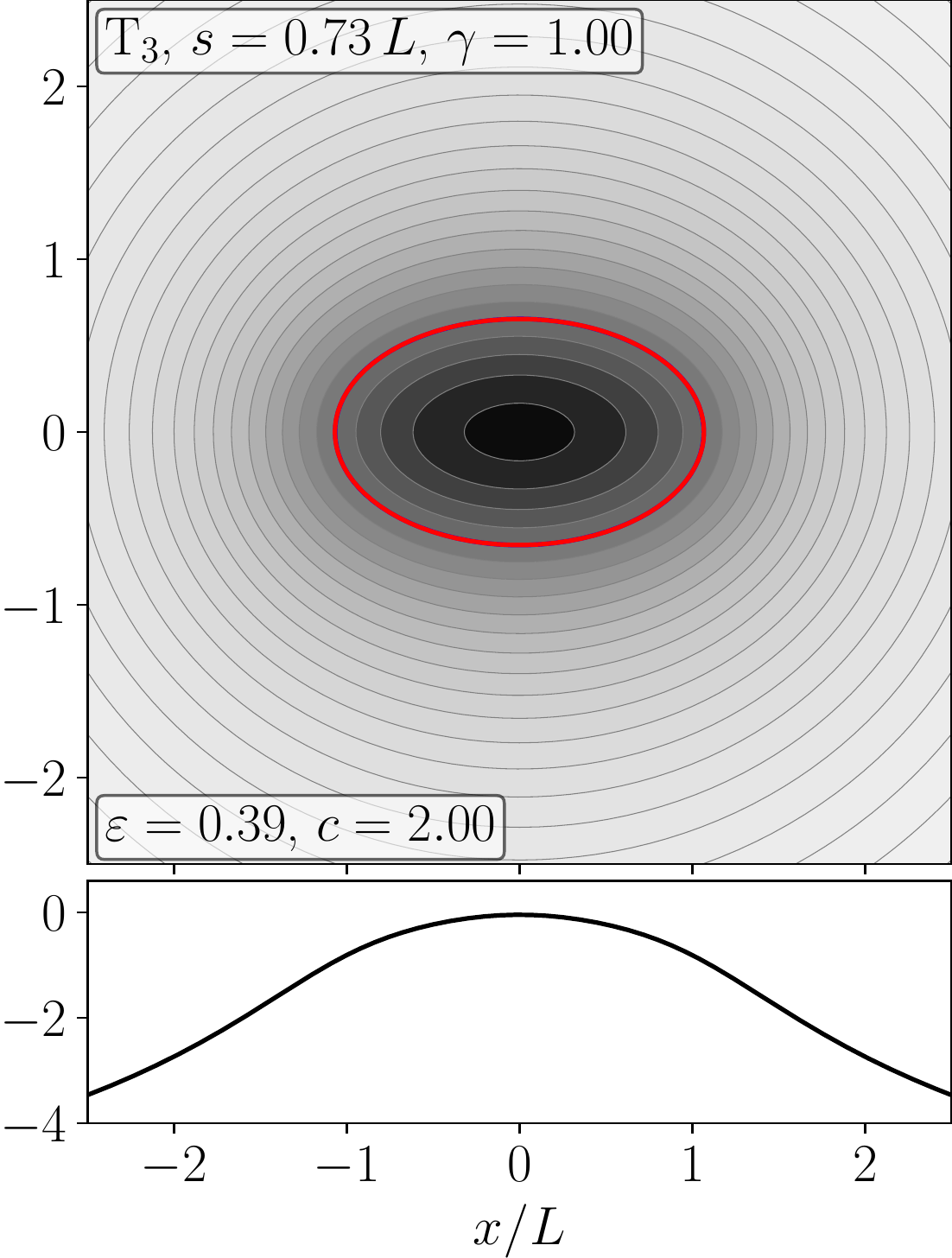}
    \end{center}
    \vspace*{-3mm}
	\caption{\label{fig:Sigma:bar:alpha}
	Like Fig.~\ref{fig:Sigma:bar:K} but for different $\gamma$ (as indicated) for model T$_3$ with $q=0.1$ and $s/L=0.73$. The first panel in row two is identical to the third in Fig.~\ref{fig:Sigma:bar:K}.
	}
\end{figure*}
%%%%%%%%%%%%%%%%%

%%%%%%%%% SECTION %%%%%%%%%%%%%%%%%%%%%%%%%%%%%%%%%%%%%
\section{Properties of single barred disc models}
\label{sec:bar:props}
We now explore the properties of individual barred models. Compound models made from several components are considered in Section~\ref{sec:compound} below.

%%%%%%%%% SUB-SECTION %%%%%%%%%
\subsection{Surface density}
We begin our exploration of these bar models in Fig.~\ref{fig:Sigma:bar:K} with the surface density of models generated from Toomre-Miyamoto-Nagai discs of different order $k=1$-4 with fixed axis ratio $q=0.1$ and $\gamma=0$, i.e.\ a flat needle-density, as used by \citetalias{LongMurali1992}. Toomre discs of increasing $k$ have ever more steeply decaying envelopes, which carries over to the barred models. One consequence of this faster radial decay is that bars constructed with the same ratio $s/L$ of scale radius to needle radius are ever thinner for increasing $k$. In Fig.~\ref{fig:Sigma:bar:K}, we have compensated this by increasing $s/L$ with $k$, such that the resulting bars have the same axis ratios.

For each model, we fit the contour with semi-major axis equal to $L$ with a generalised ellipse \citep{AthanassoulaEtAl1990}
\begin{align}
	\label{eq:general:ellipse}
	|x/a_x|^c + |y/a_y|^c = 1,
\end{align}
shown in red (mostly overlapping with the actual contour). For $c=2$ this is an exact ellipse, while $c<2$ obtains rhombic and $c>2$ box-shaped contours. Galactic bars generally have boxy contours with $c$ in the range 2.5-4 (\citeauthor{AthanassoulaEtAl1990}), rather similar to our models.

We can also see from Fig.~\ref{fig:Sigma:bar:K} that the bars have no clear edge. Instead the contours become ever rounder at radii larger than $L$. This is in contrast to real galactic bars, which are relatively sharply truncated near their co-rotation radius, outside of which galactic discs are close to axially symmetric or dominated by spiral arms. This deviation from realism implies that a single model alone cannot be used to represent a disc with a bar. However, it can still be used as a component representing the bar. For such component models, the steeper outer truncation of the higher-$k$ models is desirable in order to reduce its contribution to the outer disc.

In Fig.~\ref{fig:Sigma:bar:S}, we examine the effect of varying the ratio $s/L$ for model T$_3$. Obviously, this ratio directly affects the bar's axis ratio, but it also affects the concentration, i.e.\ how quickly in terms of $L$ the surface density reaches its asymptotic radial decay at large radii.

%%%%%%%%%%%%%%%%%
\begin{figure*}
    \begin{center}
	    \includegraphics[height=56mm]{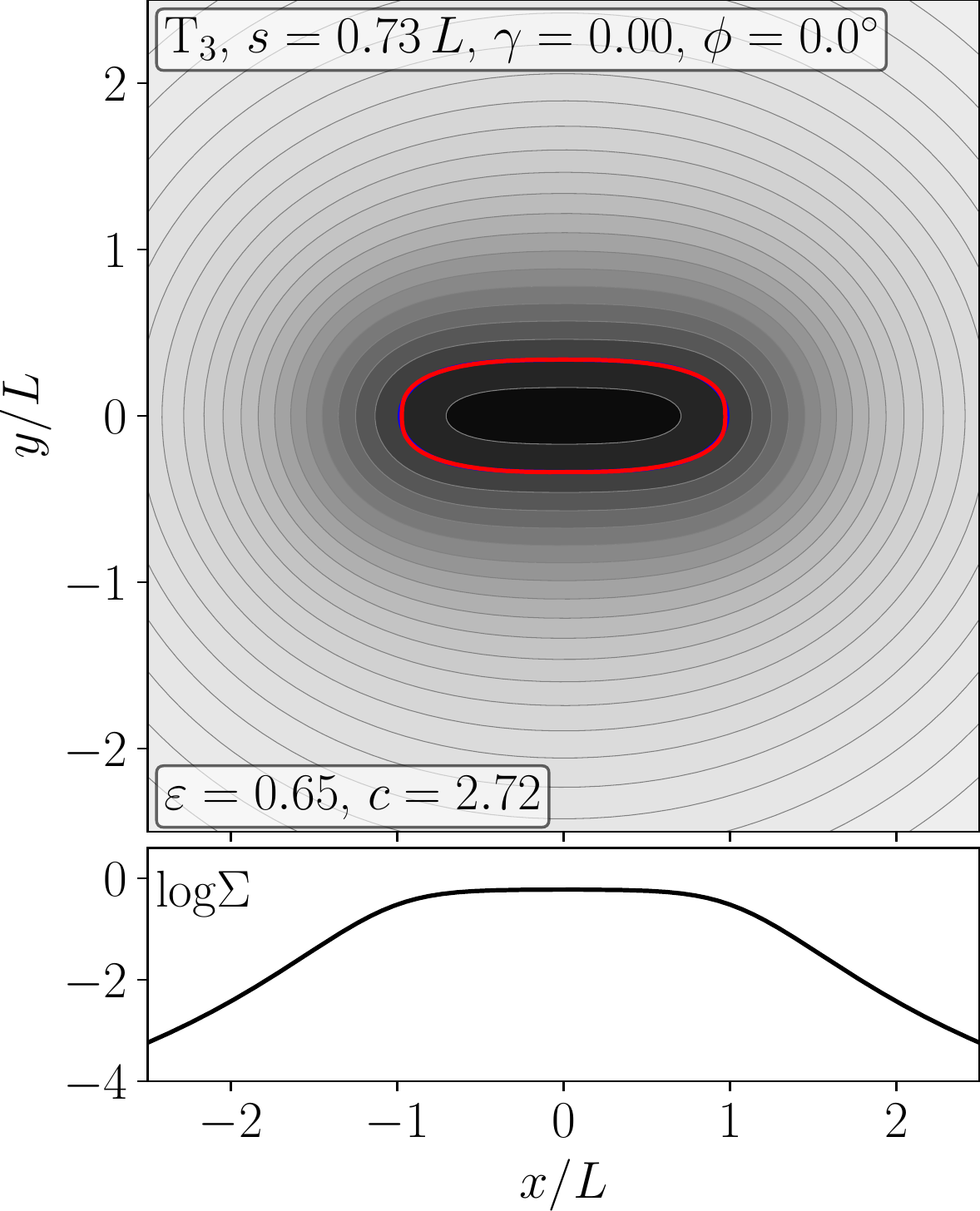}\hfil
	    \includegraphics[height=56mm]{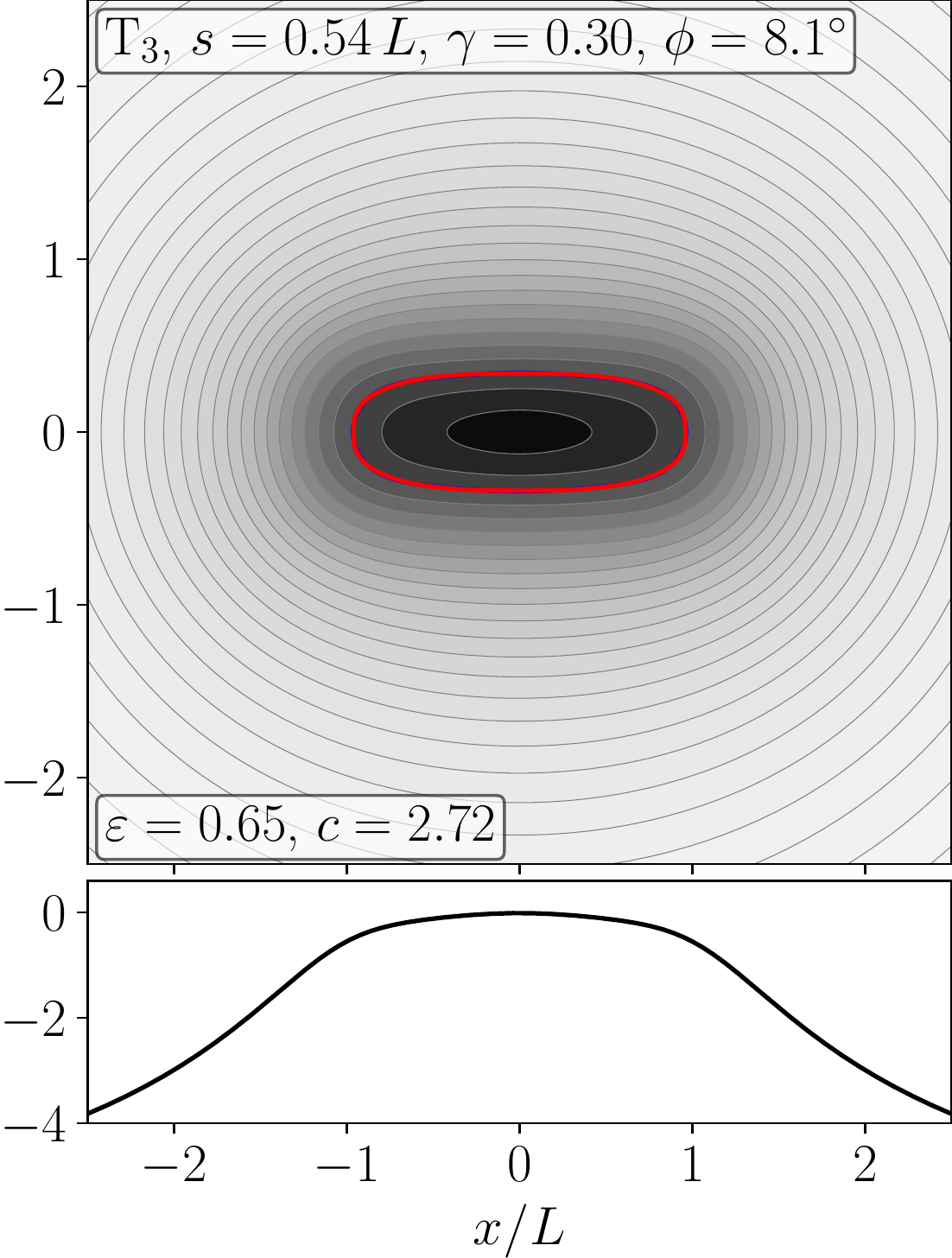}\hfil
 	    \includegraphics[height=56mm]{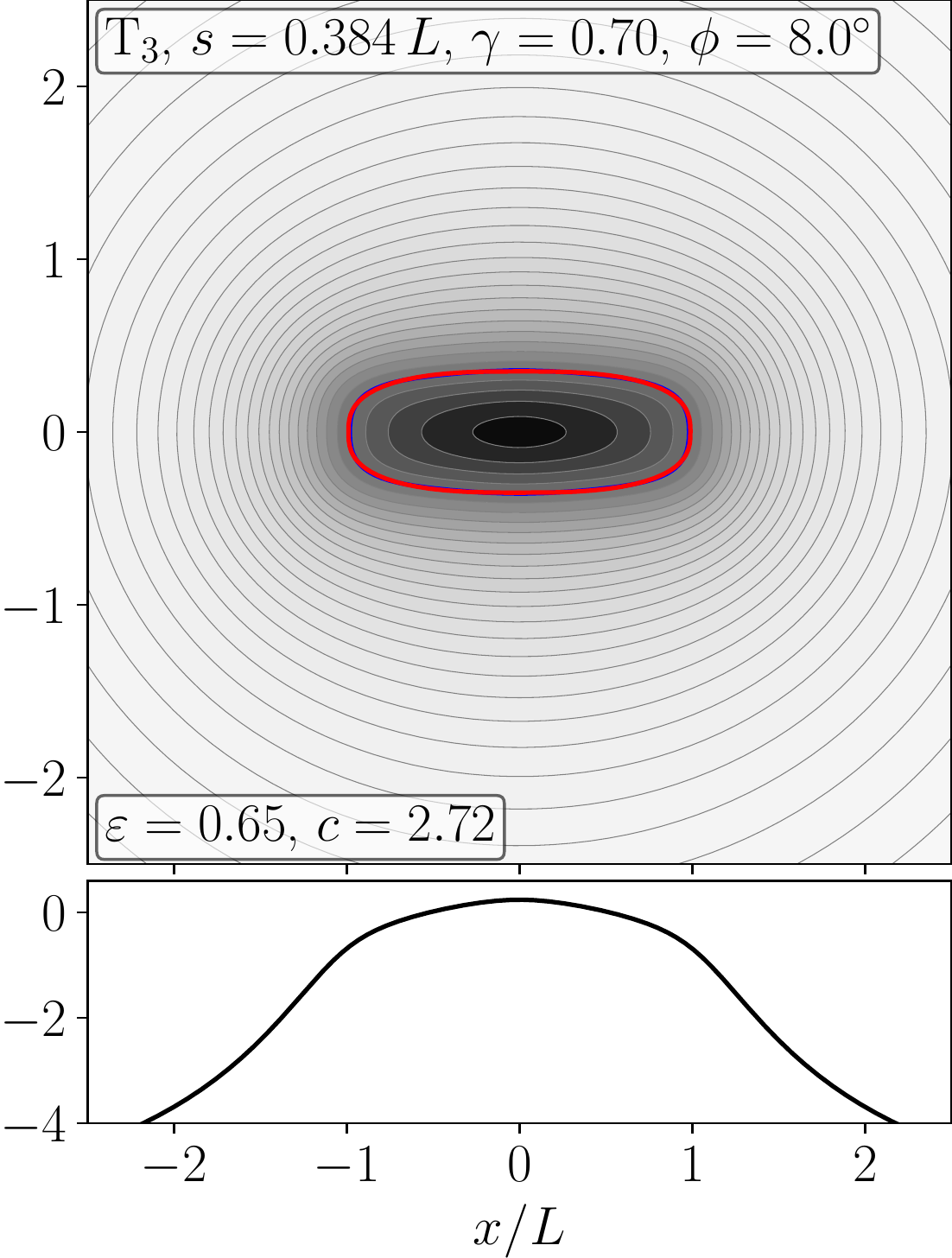}\hfil
 	    \includegraphics[height=56mm]{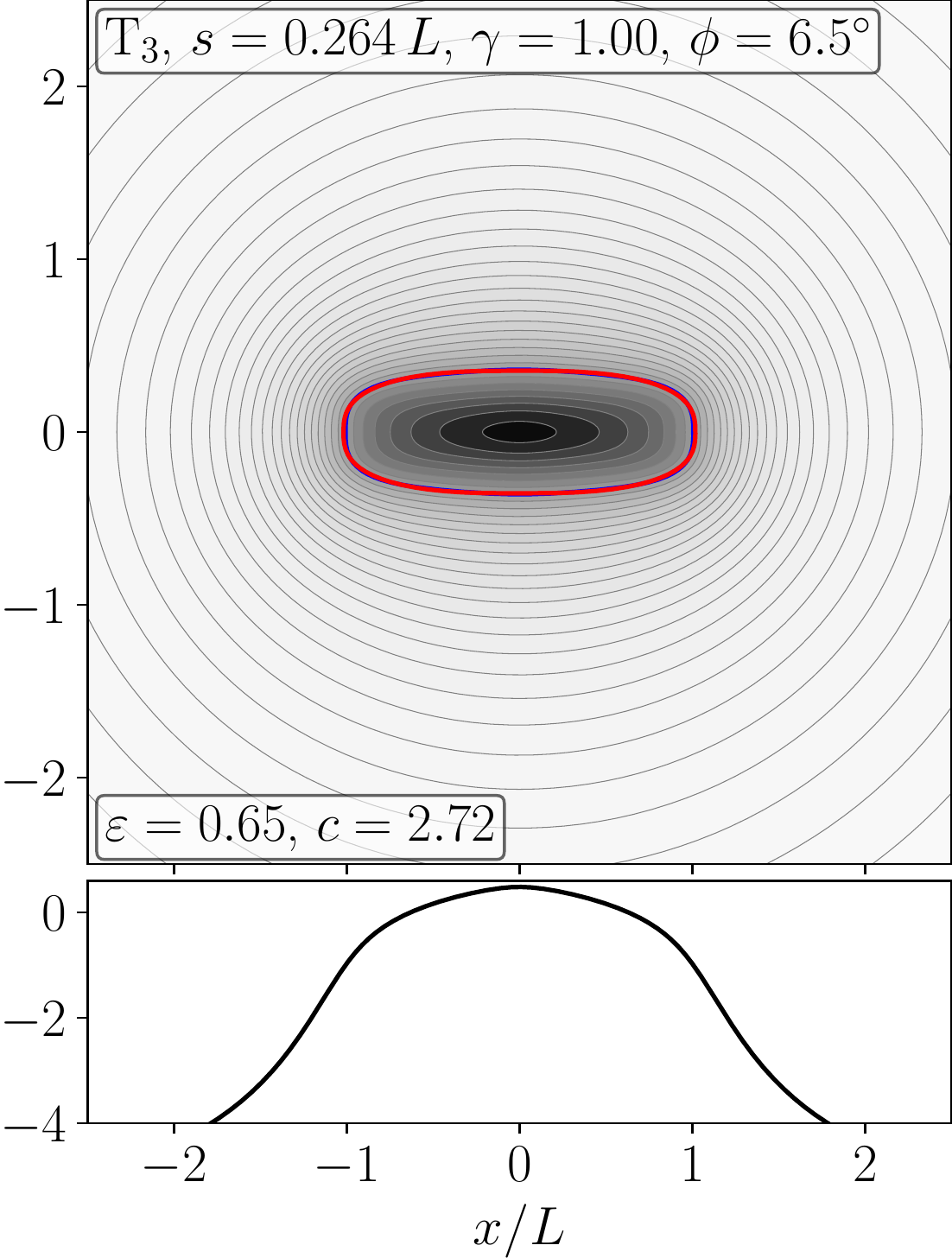}
    \end{center}
    \vspace*{-3mm}
	\caption{\label{fig:Sigma:bar:Phi}
    Surface density of T$_3$ models with the same $\gamma$ as in
	the bottom row of Fig.~\ref{fig:Sigma:bar:alpha}, but with ratios $s/L$ and angle $\phi$ (see text) adjusted to obtain similar ellipticity $\varepsilon$ and boxiness $c$.
	}
\end{figure*}
%%%%%%%%%%%%%%%%%
\begin{figure*}
    \begin{center}
        \includegraphics[height=28.5mm]{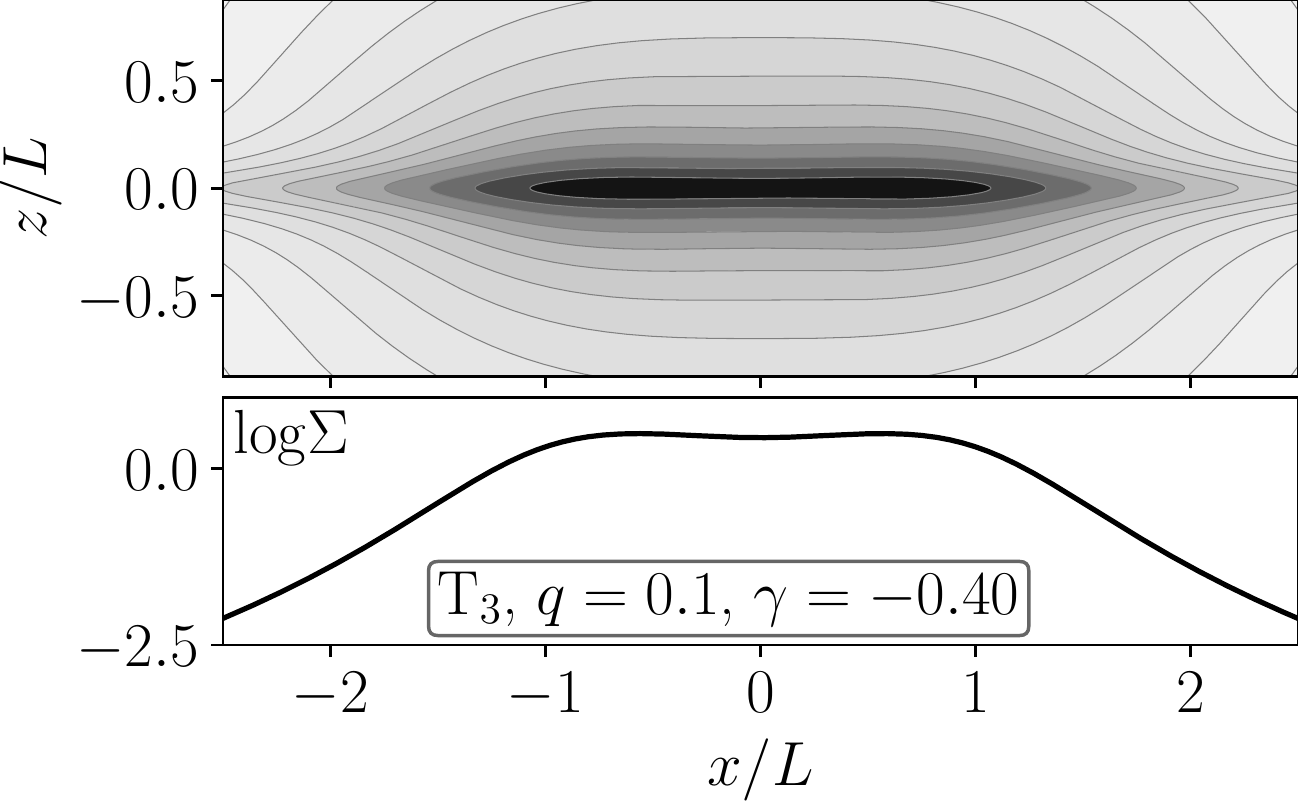}\hfil
	    \includegraphics[height=28.5mm]{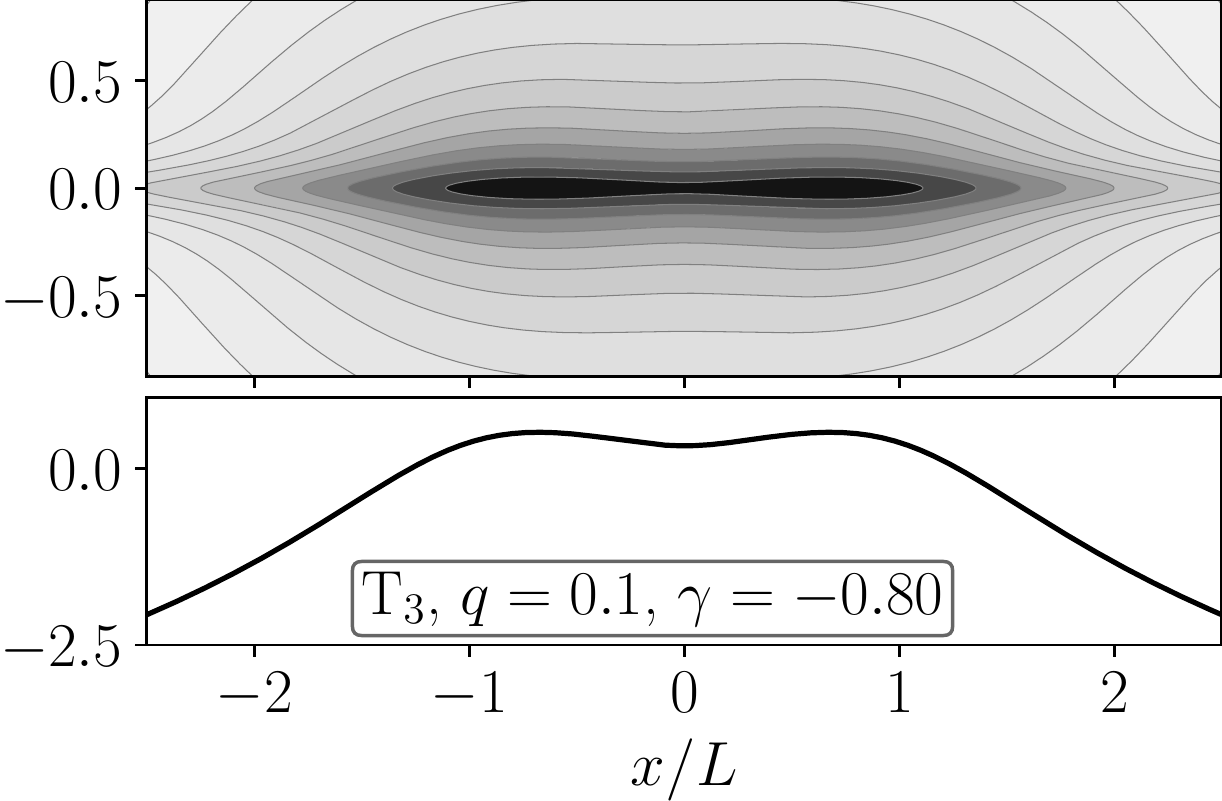}\hfil
 	    \includegraphics[height=28.5mm]{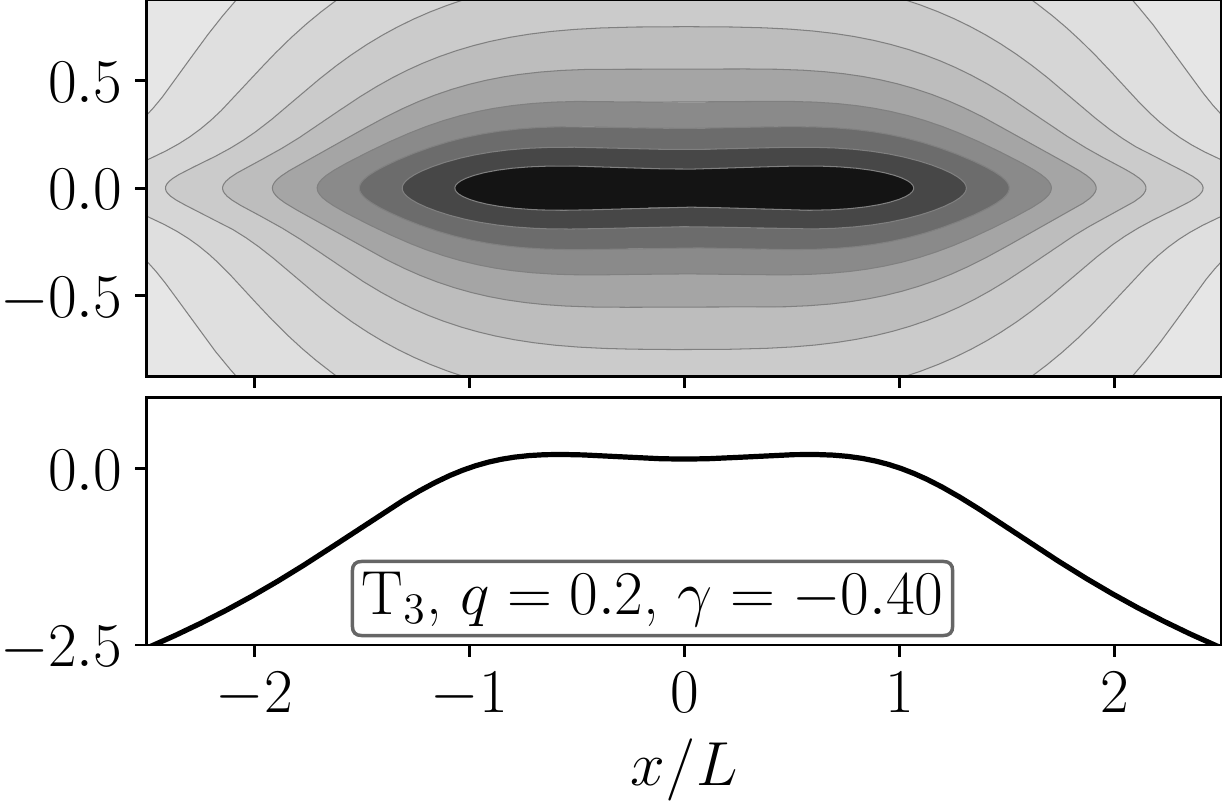}\hfil
 	    \includegraphics[height=28.5mm]{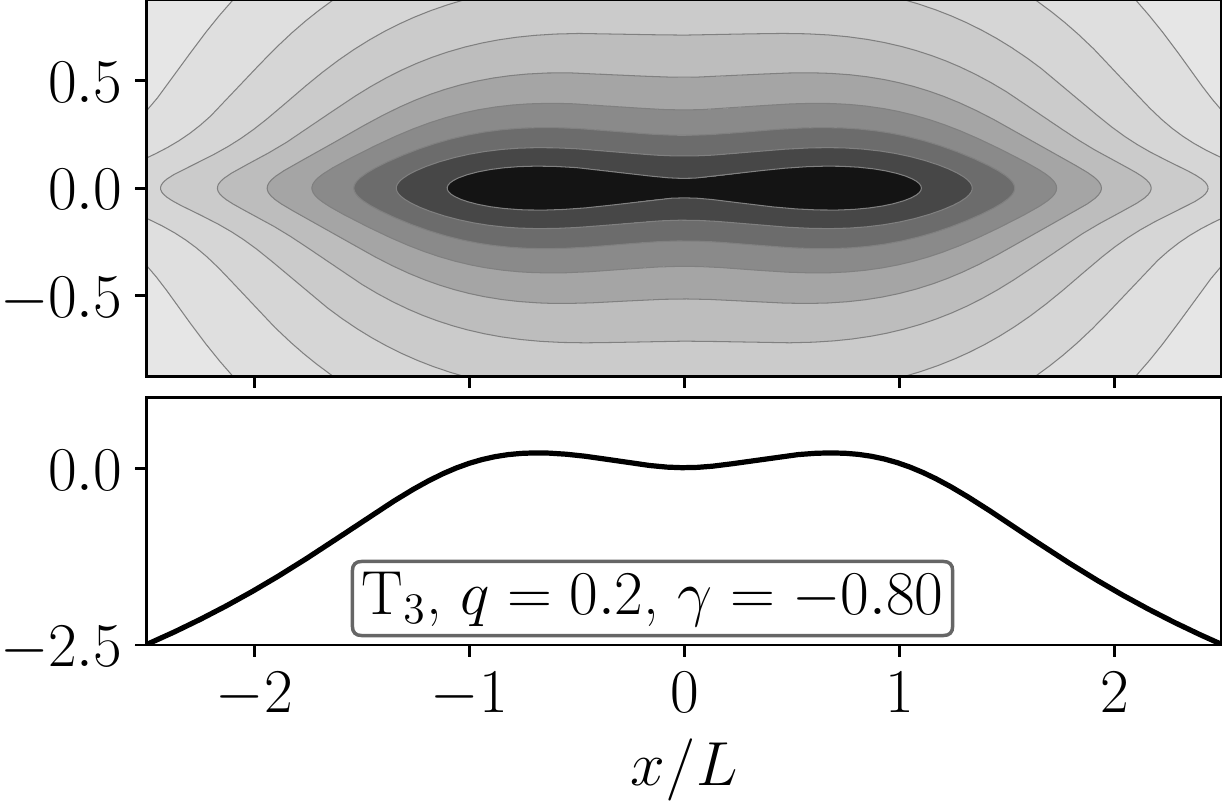}

\vspace*{2mm}

        \includegraphics[height=28.5mm]{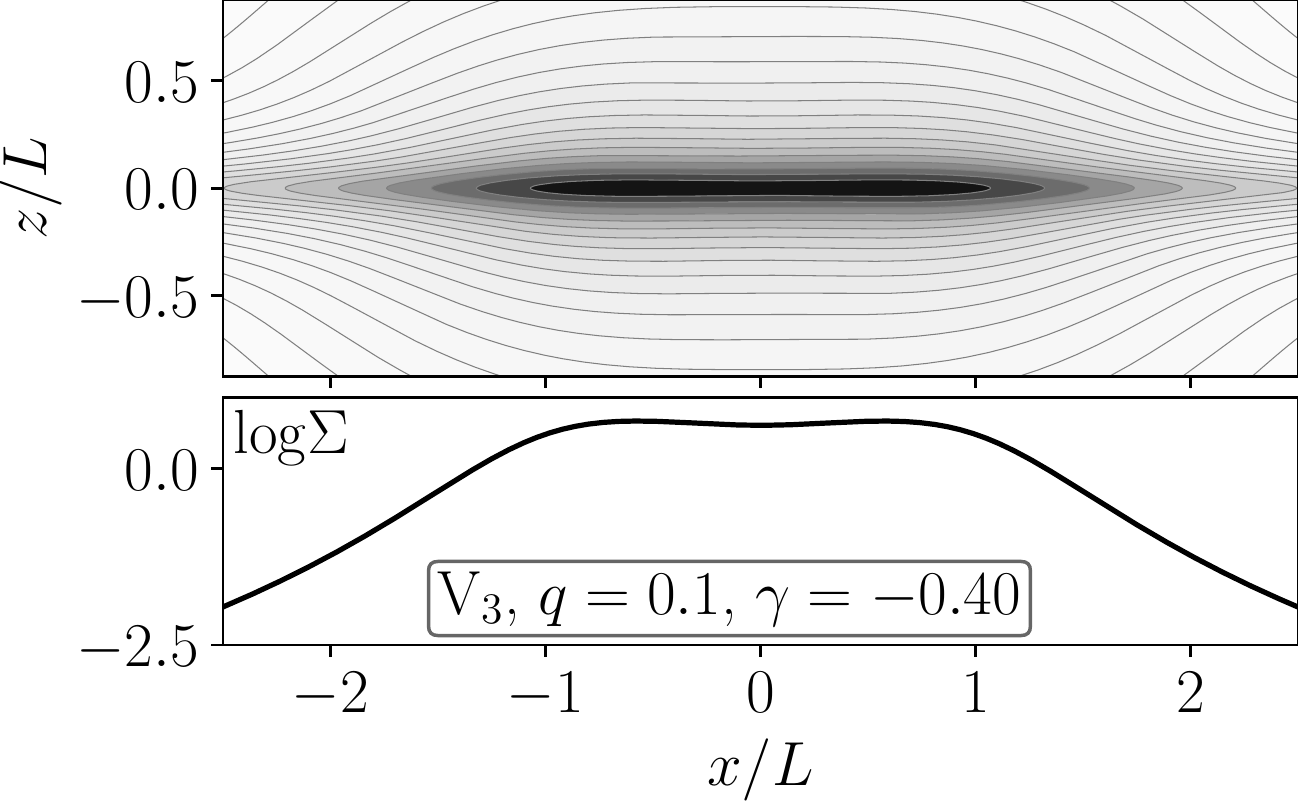}\hfil
	    \includegraphics[height=28.5mm]{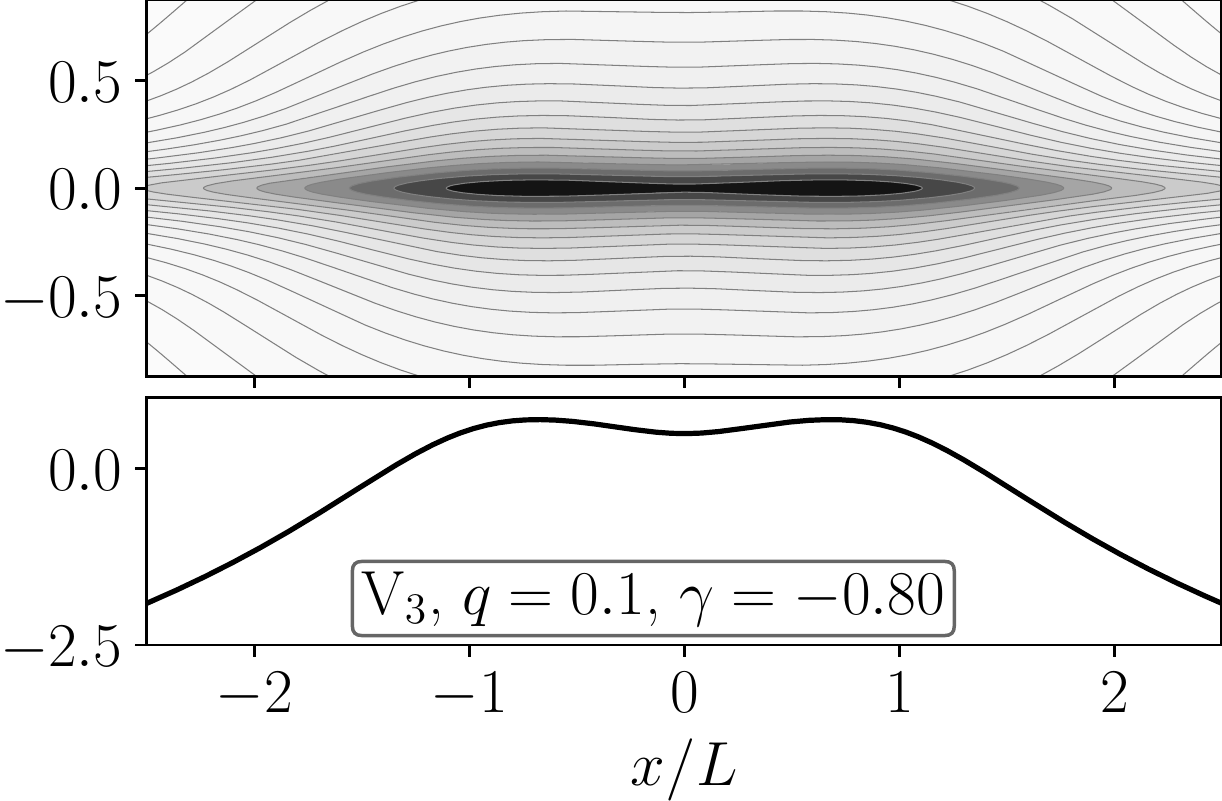}\hfil
 	    \includegraphics[height=28.5mm]{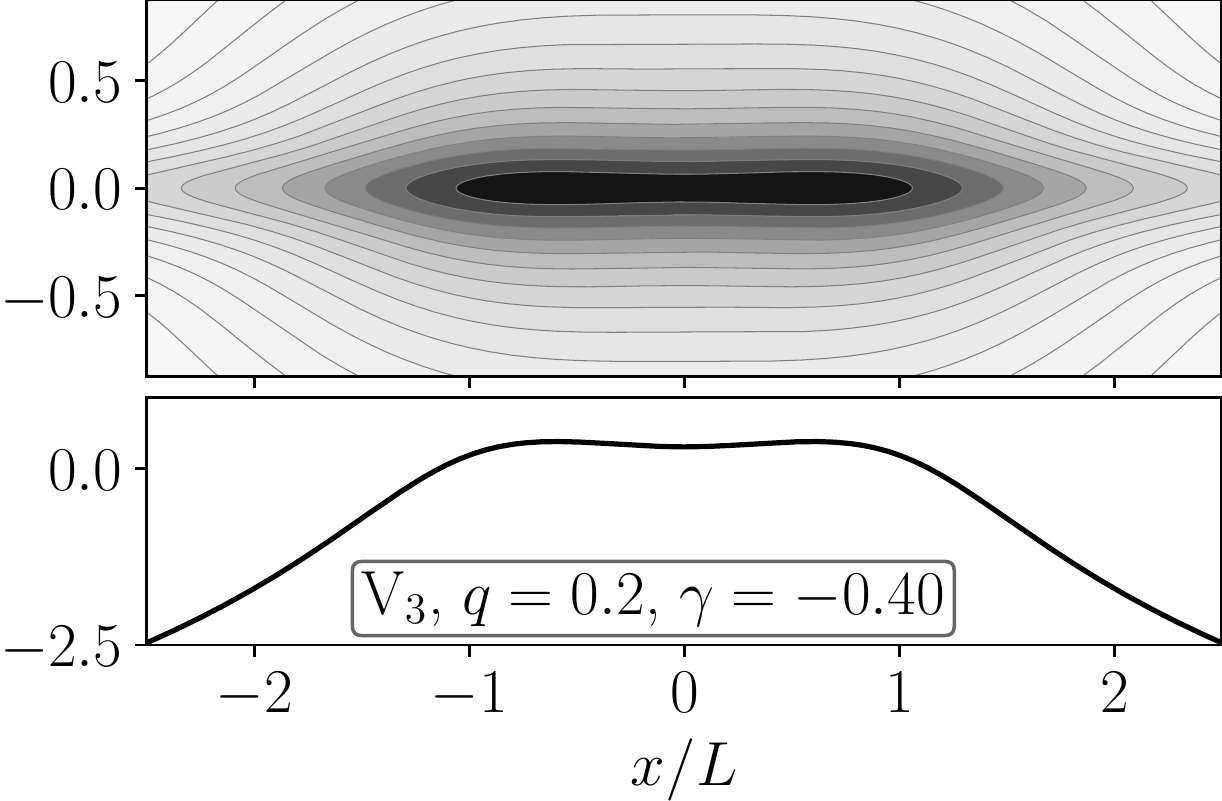}\hfil
 	    \includegraphics[height=28.5mm]{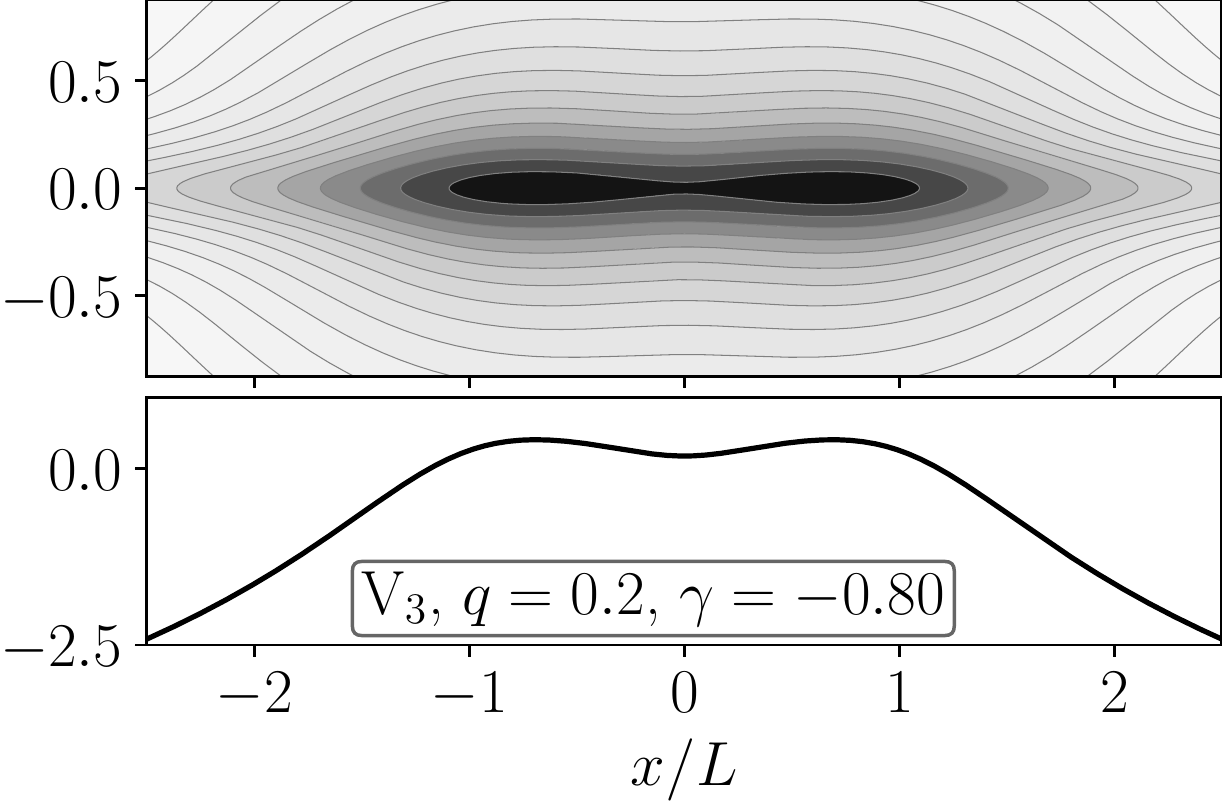}
    \end{center}
    
    \vspace*{-3mm}
	\caption{\label{fig:Peanut}
    Sideways (along $y$-axis) projected density of model T$_3$ (top) and V$_3$ (bottom) for $s=0.73L$, $q=0.1$ (left) or $q=0.2$ (right), and $\gamma=0.4$ (even panels) or $\gamma=-0.8$ (odd panels), as indicated. Contours of the spatial density in the $y=0$ plane are very similar and hence not shown. The contours above the disc are boxy for $\gamma=-0.4$ and become even dimpled (peanut-shaped) for $\gamma=-0.8$.
}
\end{figure*}
%%%%%%%%%%%%%%%%%

Fig.~\ref{fig:Sigma:bar:alpha} compares the surface density for model T$_3$ with different $\gamma$, which determines the linear gradient of the needle density. Negative values give an outwardly increasing needle density and dumbbell-like models. These are not realistic by themselves but useful as building blocks to model boxy peanuts (see below). Mildly negative $\gamma$ produces stronger boxiness in conjunction with a flatter and eventually bi-modal surface density. Positive $\gamma$ corresponds to radially decreasing needle-density and produces the opposite effect: more elliptic and more centrally concentrated surface density. The correlation with ellipticity $\varepsilon$ and boxiness $c$ along this sequence of varying $\gamma$ is similar to that observed with galactic bars \citep[e.g.][]{Gadotti2011}.

However, this correlation limits the freedom to model boxiness, central concentration, and ellipticity independently. We therefore introduce another parameter, the angle $\phi$ by which each of a pair of needles is rotated away from the $x$-axis in either direction. In other words, for $\phi\neq0$, instead of convolving the axisymmetric models with a single needle along the $x$-axis, we convolve it with two needles that cross at the origin with angle $2\phi$ between them but are otherwise identical. Fig.~\ref{fig:Sigma:bar:Phi} plots the surface density for T$_3$ models with the same values for $\gamma$ as in the bottom row of Fig.~\ref{fig:Sigma:bar:alpha}, but with $s/L$ and $\phi$ adapted such that these models have similar ellipticity $\varepsilon$ and boxiness $c$, despite their quite different radial profiles.

%%%%%%%%% SUB-SUB-SECTION %%%%%%%%%
\subsection{Boxy/peanut components}
\label{sec:BP}
In Fig.~\ref{fig:Peanut} we explore the effect of different axis ratios $q$ and negative $\gamma$ (outwardly increasing needle density) for the sideways projected density for models T$_3$ and V$_3$. These models look boxy or even dimpled like a peanut, similar to the inner boxy/peanut regions of barred galaxies. In these models, the peanut is not generated by increasing the thickness of the disc locally in the peanut region, as in galaxies, but by a bimodal component of constant thickness. In other words, these models lack a thinner inner part compared to real bars. This lack can be compensated by combining with thin models in the inner regions.

%%%%%%%%% SECTION %%%%%%%%%%%%%%%%%%%%%%%%%%%%%%%%%%%%%
\section{Compound models for barred galaxy discs}
\label{sec:compound}
%%%%%%%%%%%%%%%%%%%%%%%%%%%%%%%%%%%%%%%%%%%%%%%%%%%%%%%

\begin{figure*}
    \begin{center}
	    \includegraphics[width=88mm]{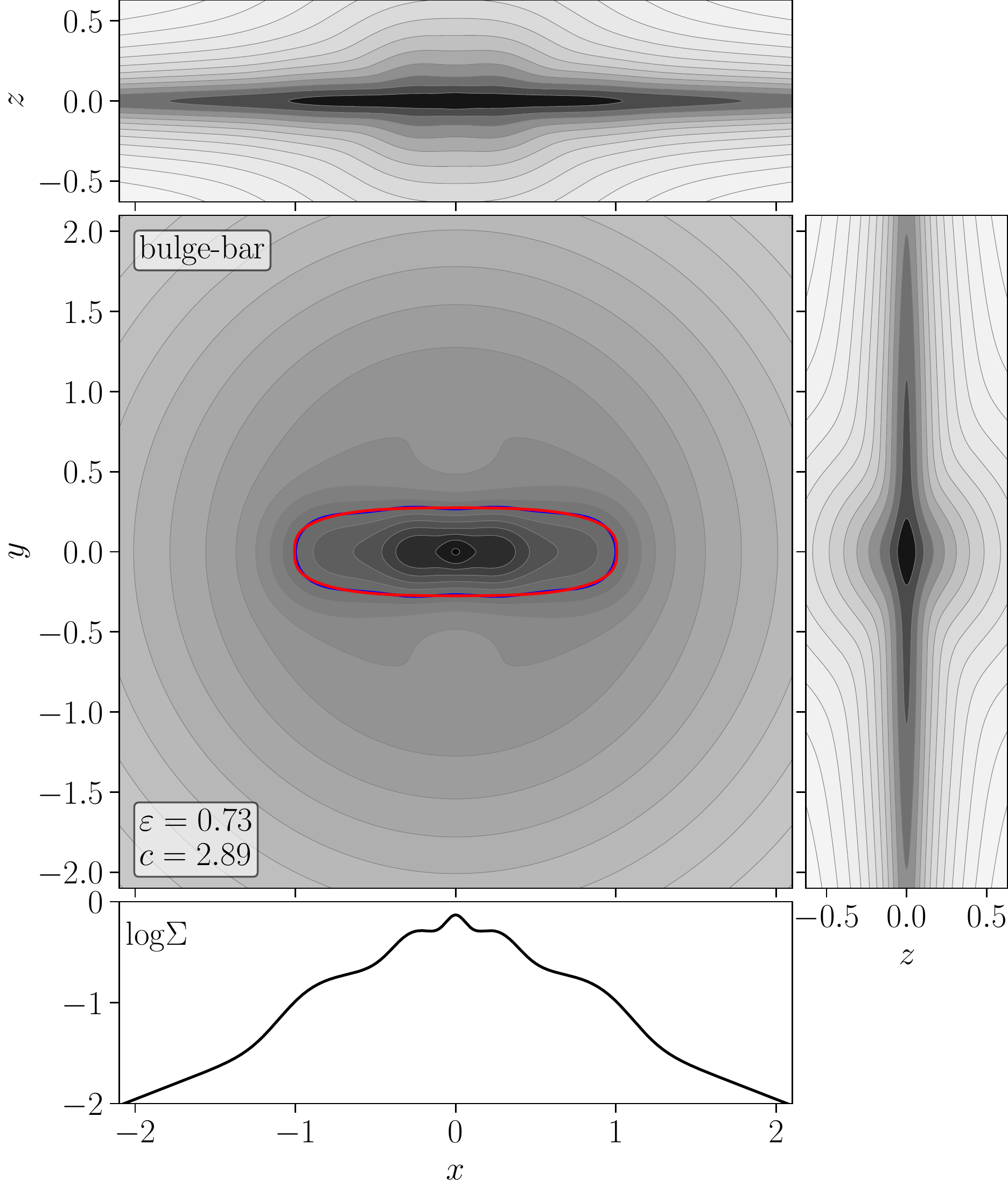}\hfill
	    \includegraphics[width=88mm]{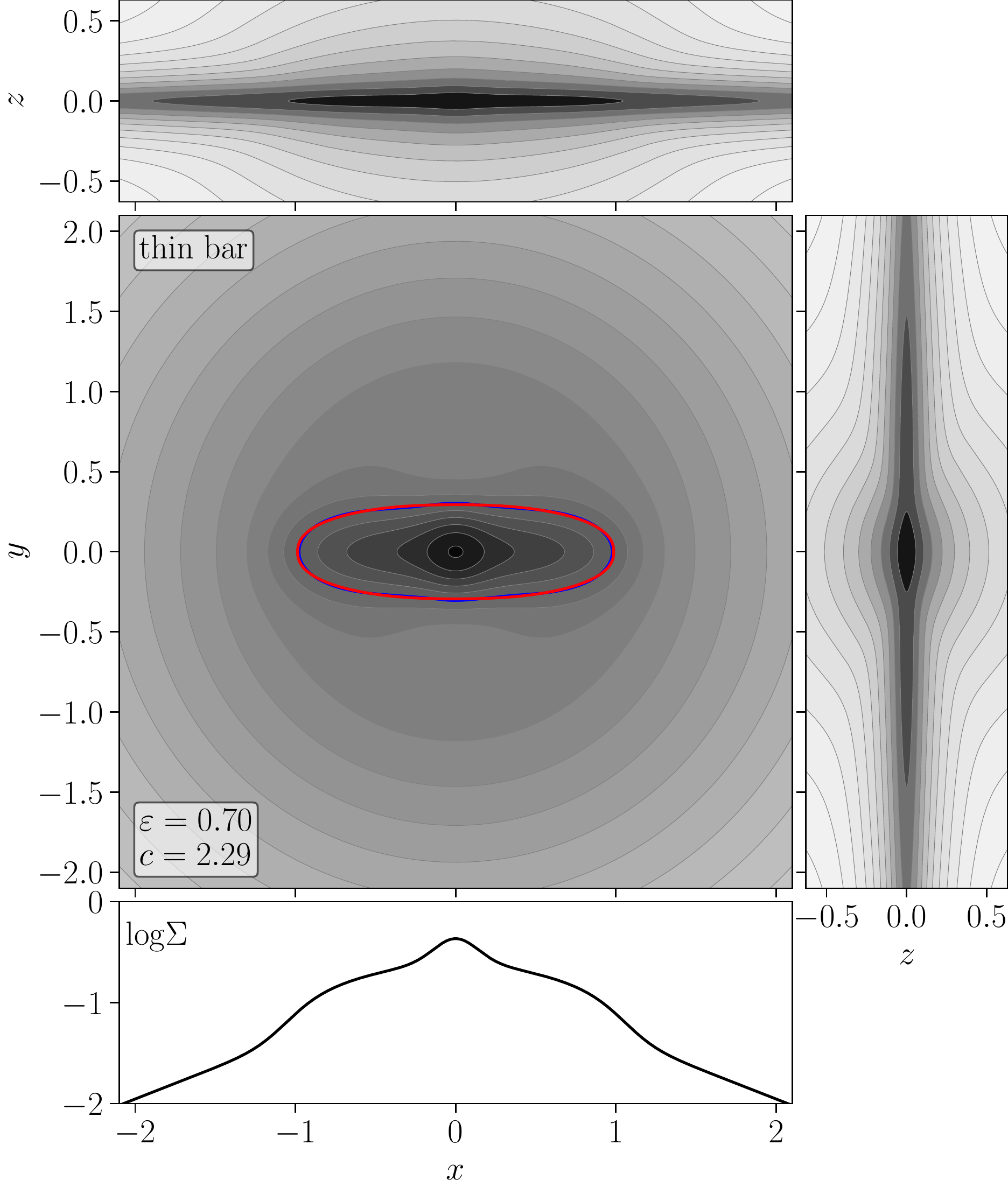}
    \end{center}
    \vspace*{-4mm}
	\caption{\label{fig:compound}
    Surface density projected along the three principal axes as well as its major-axis profile for the face-on projection (bottom) for the compound disc models `bulge-bar' (left) and `thin bar' (right) with components given in Table~\ref{tab:compound}. The ellipticity $\varepsilon$ and boxiness $c$ of the generalised ellipse (red, equation~\ref{eq:general:ellipse}) fitted to the contour with major axis equal to $L$ is given in the bottom left corner of the main plot.
	}
\end{figure*}

\begin{table}
    \caption{Parameters for the components of the compound models in Fig.~\ref{fig:compound}. \label{tab:compound} }
    \begin{tabular}{@{}lcllllll@{}}
    component    & 
    model & 
    \multicolumn{1}{c}{$M$} &
    \multicolumn{1}{c}{$a$} &
    \multicolumn{1}{c}{$b$}  &
    \multicolumn{1}{c}{$L$}  &
    \multicolumn{1}{c}{$\gamma$} &
    \multicolumn{1}{c}{$\phi$} \\
    \hline
    \multicolumn{8}{c}{bulge-bar model (left in Fig.~\ref{fig:compound})} \\
    \hline
    B/P bulge    & V$_4$ & $\phm0.08$ & 0.05 & 0.25 & 0.33 & $-0.95  $ & $0\degr$ \\
    bar          & V$_4$ & $\phm0.15$ & 0.4  & 0.1  & 1    & $\phm0.1$ & $6\degr$ \\
    nuclear disc & V$_4$ & $\phm0.01$ & 0.1  & 0.1  & 0    & - & - \\
    main disc    & V$_2$ & $\phm1 $   & 1.52 & 0.08 & 0    & - & - \\
    hole in disc & V$_2$ & $-0.545$   & 1.12 & 0.08 & 0    & - & -  \\
    \hline 
    \multicolumn{8}{c}{thin bar model (right in Fig.~\ref{fig:compound})} \\
    \hline
    bar          & T$_3$ & $\phm0.15$ & 0.3  & 0.1  & 1    & $\phm0.7$ & $0\degr$ \\
    nuclear disc & V$_4$ & $\phm0.02$ & 0.3  & 0.1  & 0    & - & -  \\
    main disc    & V$_2$ & $\phm1 $   & 1.52 & 0.08 & 0    & - & -  \\
    hole in disc & V$_2$ & $-0.545$   & 1.12 & 0.08 & 0    & - & -
    \end{tabular}
\end{table}

In order to build realistic models for barred galaxy discs, we now combine axisymmetric with barred disc models. To this end, we first construct an axisymmetric disc with a central hole, into which the bar can be placed. We do this by subtracting two disc models of the same type and with the same scale height $b$ and central density $\rho(0)$, but with different scale lengths $a$ and masses $M$.

Next, we add the bar component, a nuclear disc, and optionally a boxy/peanut bulge. For these central components to be negligible outside the bar region, their model types should have higher order $k$ than those used for the outer disc, such that their radial decay is steeper. 

In this way, we constructed by trial and error two illustrative disc models, whose components are detailed in Table~\ref{tab:compound} and whose surface densities projected along the three principal axes are shown in Fig.~\ref{fig:compound}. The hollow disc component is the same for both models. The `bulge-bar' model has a boxy/peanut bulge component, while the `thin bar' model does not. 

The surface density projections along the bar major axis in Fig.~\ref{fig:compound} (bottom panels) have a flatter profile in the bar region ($|x|\lesssim1$ than further out. This `flat' profile is typical of stronger bars \citep{ElmegreenElmegreen1985} and also known as `shoulder' feature. Recently, \cite{AndersonEtAl2022} presented evidence through numerical simulations that this feature is an indication of bar secular growth, and is often accompanied by a boxy/peanut bulge, which itself is present in our bulge-bar model in Fig.~\ref{fig:compound}. As can be seen in Figs.~\ref{fig:Sigma:bar:K}-\ref{fig:Sigma:bar:Phi}, most of our individual models have such flat profiles, which is a consequence of the way they are constructed by convolution with a near-constant density needle. In order to construct models with less flat bar profiles, typical for weaker bars, one can try $\gamma\sim1$ and $s\ll L$ or must add more centrally concentrated components.

\begin{figure}
    \includegraphics[width=\columnwidth]{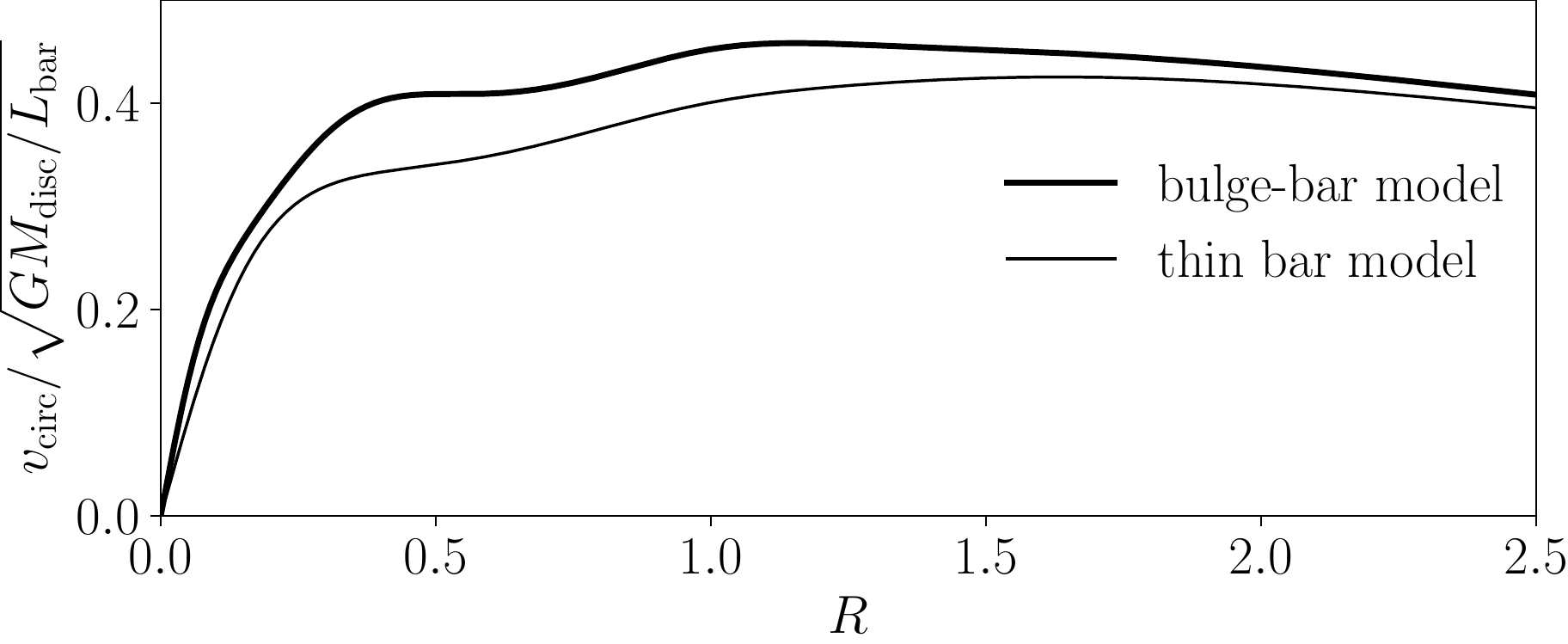}
    \vspace*{-5mm}
	\caption{\label{fig:compound:vcirc}
    The circular-speed curves computed from the axisymmetric part of the compound disc models of Table~\ref{tab:compound} (see also Fig.~\ref{fig:compound}).}
\end{figure}

\begin{figure}
    \includegraphics[width=\columnwidth]{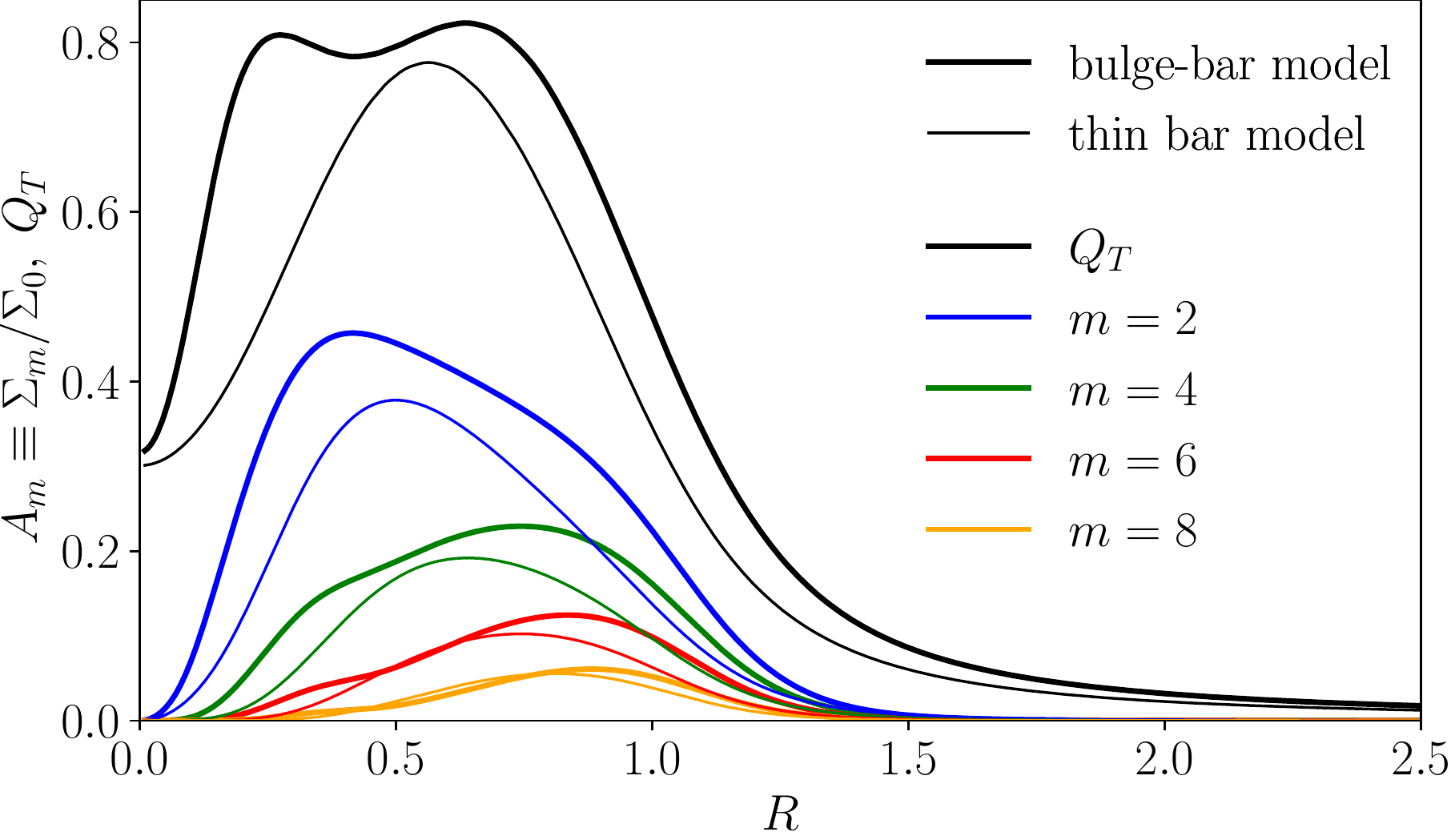}
    \vspace*{-5mm}
	\caption{\label{fig:compound:strength}
    The measure $Q_T$ defined in equation~\eqref{eq:QT} for the relative strength of the non-axisymmetric forces and the relative Fourier amplitudes of the surface density for the compound models of Table~\ref{tab:compound} (see also Fig.~\ref{fig:compound}).}
\end{figure}

In Fig.~\ref{fig:compound:vcirc}, we plot the circular-speed curves of these discs, which lack the typical central peak from a classical bulge. Fig.~\ref{fig:compound:strength} shows the relative Fourier amplitudes $A_m\equiv\Sigma_m/\Sigma_0$ of the surface density. The `thin bar' model is rhombic inside and mildly boxy further out, resulting in the maximal $A_m$ at larger $R$ for larger $m$. The `bulge-bar' model has two barred components, a wide range of isophote shapes and more complex $A_m$ profiles. In particular, the strong outer boxiness manifests as $A_{m>2}$ peaking near the end of the bar.

Fig.~\ref{fig:compound:strength} also plots in black the ratio
\begin{align}
    \label{eq:QT}
    Q_T(r) = \frac{\max_\varphi\{|\p \Phi/\p\varphi|\}}{r \langle|\p \Phi/\p r|\rangle_\varphi}
\end{align}
between the maximum tangential and average radial force at any given radius in the equatorial plane ($z=0$), proposed as measure of bar strength \citep{CombesSanders1981}. Any spheroidal components not included in our model, such as a dark halo or classical bulge, would contribute to the radial but not to the tangential forces and hence reduce $Q_T$ below the values found here. Most likely, this is the reason why our $Q_T$ values in Fig.~\ref{fig:compound:strength} are somewhat larger than those $Q_T\lesssim0.6$ inferred for observed bars \citep{ButaBlock2001, LaurikainenEtAl2004, SaloEtAl2010}

We see that the transition from the barred to the round region is quite sharp for these compound models, in agreement with real and simulated bars and in contrast to the single models of Section~\ref{sec:bar:props}. 

%%%%%%%%% SECTION %%%%%%%%%
\section{Summary}
\label{sec:conclude}
In Section~\ref{sec:bar}, we extend the triaxial bar model of \cite{LongMurali1992}, which is a convolution of the axisymmetric \cite{MiyamotoNagai1975} disc with a box function $f(x)$,  in three ways. First, we replace the box function by a linear function of $|x|$ in the range $|x|<L$. This enables bar models whose density along the bar is not flat, but decreasing or even increasing. The latter is useful for making boxy/peanut components. Secondly, we consider various axisymmetric input models, which have steeper vertical and radial decay at large radii than the Miyamoto-Nagai disc. Such steeper profiles are useful when constructing compound models, as they reduce unwanted contributions of the central components to the outer regions. Finally, we consider two such models, twisted by $\pm\phi$ away form the $x$-axis. This enables us to model the boxiness and ellipticity of the density contours independently of the central concentration.

The surface densities of these barred models are shown and explored in Section~\ref{sec:bar:props}. Individual models have already some striking similarities to galactic bars, such as the correlation between the ellipticity and boxiness of their surface-density contours and flat major-axis profiles for strongly barred models. However, there are also some differences. First, our models tend to continuously become rounder at radii larger than the radial extend $L$ of the needle used in the construction of the models, while real galaxies change rather abruptly from barred to round. Second, unlike galactic bars, the models cannot have features like ansea at the ends of the bars or deviations from triaxial symmetry (e.g.\ asymmetric ansea).

Some of these deficiencies can be overcome by combining several individual models, as we demonstrate with two simple examples in Section~\ref{sec:compound}. In particular, the compound models can have an abrupt end to the bar and an inner vertically thicker boxy/peanut component. In this work, we refrain from a thorough exploration of the possibilities of combining our models, including with models of different type for example exponential discs, but it is quite obvious that with these novel bar models galactic bars can be modelled with analytical potentials in a much more realistic way than possible so far.

%%%%%%%%%%%%%%%%%%%%%%%%%%%%%%%%%%%%%%%%%%%%%%%%%%%%%%%
\section*{Acknowledgements}
%%%%%%%%%%%%%%%%%%%%%%%%%%%%%%%%%%%%%%%%%%%%%%%%%%%%%%%
We thank Ralph Schönrich, Lia Athanssoula, and Marcin Semczuk for many stimulating conversations, Peter Erwin for an informative discussion, the first reviewer, James Binney, for a prompt and valuable report, and Jo Bovy for pointing us to the study by \citeauthor{LongMurali1992} before it was too late. This work was partially funded by the Deutsche Forschungsgemeinschaft (DFG, German Research Foundation) -- SFB 881 (``The Milky Way System'', sub-project P03).

%%%%%%%%%%%%%%%%%%%%%%%%%%%%%%%%%%%%%%%%%%%%%%%%%%%%%%%
\section*{Data Availability}
%%%%%%%%%%%%%%%%%%%%%%%%%%%%%%%%%%%%%%%%%%%%%%%%%%%%%%%
Computer codes in \texttt{python} and \texttt{C++} for the computation of gravitational potential, forces, force gradients (\texttt{C++} only), density, and projected density (\texttt{python} only) are available at \url{https://github.com/WalterDehnen/discBar}.

%%%%%%%%%%%%%%%%%%%%%%%%%%%%%%%%%%%%%%%%%%%%%%%%%%%%%%%
\bibliographystyle{mnras}
\bibliography{discBar}
%%%%%%%%%%%%%%%%%%%%%%%%%%%%%%%%%%%%%%%%%%%%%%%%%%%%%%%
\onecolumn \appendix
%%%%%%%%%%%%%%%%%%%%%%%%%%%%%%%%%%%%%%%%%%%%%%%%%%%%%%%
\section{Convolution integrals}
\label{app:In}
%%%%%%%%%%%%%%%%%%%%%%%%%%%%%%%%%%%%%%%%%%%%%%%%%%%%%%%
In this appendix, we derive closed forms for the convolutions integrals~(see equation~\ref{eq:In})
\begin{align}
	\label{eq:conv:def}
    \mathcal{I}_n(\vec{r};L,\gamma) %= \sub{\rho}{needle}\ast\frac1{r^n}
    = \frac1{2L}\int_{x-L}^{x+L} \left(1+\gamma-2\gamma\frac{|x-t|}{L}\right)\,\frac{\diff t}{(t^2+u^2)^{n/2}},
\end{align}
where $u\equiv\sqrt{y^2+z^2}$. Given the functions
\begin{align}
	A_1(\vec{r}) = \ln (r+x),\quad
	A_2(\vec{r}) = \frac1{u}\arctan\frac{x}u,\quad
	A_{n+2}(\vec{r}) = \frac1{nu^2}\left[(n-1) A_n(\vec{r}) + \frac{x}{r^n}\right],
    \quad\text{and}\quad
	B_n(\vec{r}) = \int \frac{\diff r}{r^{n-1}},
\end{align}
which satisfy $\p A_n/\p x = 1/r^n$ and $\p B_n/\p x = x/r^n$, we have
\begin{align}
	\label{eq:conv}
    \mathcal{I}_n(\vec{r};L,\gamma)
	&= \frac{1-\gamma}{2L} \big[A_n(\vec{r}+L\uvec{x})-A_n(\vec{r}-L\uvec{x})\big] + \frac{\gamma}{L^2}\big[C_n(\vec{r}+L\uvec{x})+C_n(\vec{r}-L\uvec{x})-2C_n(\vec{r})\big],
\end{align}
where $C_n(\vec{r}) \equiv xA_n(\vec{r})-B_n(\vec{r})$. Since $\p C_n/\p x=A_n$, $\p I_{n}/\p x$ is given by equation~\eqref{eq:conv} after the replacements $A_n\to1/r^n$ and $C_n\to A_n$, while $\p\mathcal{I}_n/\p y=-n y\mathcal{I}_{n+2}$ and $\p\mathcal{I}_n/\p z=-n z \mathcal{I}_{n+2}$.
%%%%%%%%%%%%%%%%%%%%%%%%%%%%%%%%%%%%%%%%%%%%%%%%%%%%%%%
\section{Axisymmetric models}
\label{app:axi}
%%%%%%%%%%%%%%%%%%%%%%%%%%%%%%%%%%%%%%%%%%%%%%%%%%%%%%%
In this appendix, we derive the axisymmetric models T$_k$ and V$_{\!k}$ discussed in Section~\ref{sec:more:models}. First, we note that given an \emph{input model} $\Psi(\vec{r})$ satisfying $\vec{\nabla}^2\Psi=0$ for $|z|>a+b$, a disc model can be constructed via the coordinate replacement \citep[see also \S2.3.1 of][]{BinneyTremaine2008}
\begin{align}
    \label{eq:recipe:MN}
    \Phi(\vec{r}) = \Psi(\vec{\varrho})
\end{align}
with $\vec{\varrho}$ given in equation~\eqref{eq:varrho}. Poisson's equation then tells us that the associated density is 
\begin{align}
	\label{eq:density:MN}
	\rho(\vec{r}) =
    \frac{b^2}{4\upi G\zeta^3} \left[\partial_Z\Psi(\vec{\varrho}) - \zeta \partial_Z^2\Psi(\vec{\varrho})\right]
    \quad \text{for $b>0$}
    \qquad \text{and} \quad \rho(\vec{r}) =
    \frac{\delta(z)}{2\upi G} \,\p_Z\Psi|_{\{x,y,a\}}
    \quad \text{for $b=0$},
\end{align}
which for the point-mass potential $\Psi=-GM/r$ as input model and $b>0$ gives the density~\eqref{eq:rho:T1} of the \citetalias{MiyamotoNagai1975} disc.

Secondly, we introduce the building blocks $\theta_n(\vec{r}) \equiv (-\partial_z)^n r^{-1}$ for the input model $\Psi(\vec{r})$. These building blocks are proportional to the spherical harmonics $Y_n^0$ and satisfy $\vec{\nabla}^2\theta=0$ at $r>0$, such that linear combinations of them (with coefficients independent of $\vec{r}$) do so too. Moreover, they allow a straightforward algebraic manipulation of the $z$ derivatives, which are ubiquitous in the relations for obtaining the various disc models below. These building blocks follow the recursion relation $r^2 \theta_{n+1} = (2n+1) z \theta_n - n^2 \theta_{n-1}$ and in terms of coordinates are given by
\begin{align}
	%\theta_{-1} &= -\ln(r+z),&
	\theta_0 &= \frac1r,&
	\theta_1 &= \frac{z}{r^3},&
	\theta_2 &= \frac{3z^2}{r^5} - \frac1{r^3},&
	%\nonumber \\ \label{eqs:psi:n}
	\theta_3 &= 3z\left[\frac{5z^2}{r^7} - \frac{3}{r^5}\right],&
	\nonumber \\ \label{eqs:psi:n}
	\theta_4 &= 3\left[\frac{35z^4}{r^9} - \frac{30z^2}{r^7} + \frac3{r^5}\right],&
	\theta_5 &= 15z\left[\frac{63z^4}{r^{11}}-\frac{70z^2}{r^9}+\frac{15}{r^7}\right],&
	\theta_6 &= 45\left[\frac{231z^6}{r^{13}}-\frac{315z^4}{r^{11}}+\frac{105z^2}{r^9}-\frac{5}{r^7}\right].
\end{align}

Next, we consider the manipulations to generate the models T$_k$ and V$_k$. To that end, we introduce the differential operator
\begin{align}
	\hat{D}_{p,n} \equiv -\frac{p^{n+1}}{n} \pdiff{}{p} \frac1{p^n}
	= 1 - \frac{p}{n} \pdiff{}{p},
\end{align}
where $p$ is some parameter ($a$ or $b$) of the models and $n$ an integer. The surface densities~\eqref{eq:Toomre} of Toomre's models follow the relation
\begin{subequations}
\begin{align}
	\Sigma_{\mathrm{T}_{k+1}}(R;a) &= \hat{D}_{a,2k-1} \cdot \Sigma_{\mathrm{T}_k}(R;a).
\end{align}
Hence, by analogy, so do potential and density (also for the finite-thickness Toomre-Miyamoto-Nagai models):
\begin{align}
	\label{eq:recur:a:phi}
	\Phi_{\mathrm{X}_{k+1}}(\vec{r};a,b) &= \hat{D}_{a,2k-1} \cdot \Phi_{\mathrm{X}_k}(\vec{r};a,b)
	\qquad\text{and}\\
	\label{eq:recur:a:rho}
	\rho_{\mathrm{X}_{k+1}}(\vec{r};a,b) &= \hat{D}_{a,2k-1} \cdot \rho_{\mathrm{X}_k}(\vec{r};a,b).
\end{align}
Here, `X' stands for `T' or `V', because the models V$_{\!k}$ follow the same relations, since the derivative $\p/\p a$ to increment $k$ and the manipulation to obtain model V$_{\!k}$ from T$_k$ can be applied interchangeably. The derivative in equation~\eqref{eq:recur:a:phi} is applied at fixed $\vec{r}$, such that when translating this relation to the input models $\Psi(\vec{r})$ one must account for the dependence of $\vec{\varrho}$ on $a$ via $Z=\zeta+a$. This gives the relation
\begin{align}
	\label{eq:recur:a}
	\Psi_{\mathrm{X}_{k+1}}(\vec{r},a)   &= \left(1-\frac{a}{2k-1}\left[\pdiff{}{a}+\pdiff{}{z}\right]\right) \cdot \Psi_{\mathrm{X}_k}(\vec{r},a)
\end{align}
\end{subequations}
for the input models. We can also define an input model $\mathcal{R}(\vec{r})$ for the density implicitly by writing
\begin{align}
	\rho_{\mathrm{X}_k}(\vec{r}) =\frac{Mb^{2+2l}}{2^{2+l}\upi\zeta^{3+2l}}
	\mathcal{R}_{\mathrm{X}_k}(\vec{\varrho}),
\end{align}
where $l$ is the number of differentiations w.r.t.\ $b$, i.e. $l=0$ for models T$_k$ and $l=1$ for models V$_{\!k}$.

%%%%%%%%%%%%%%%%%%%%%%%%%%%%%%%%%%%%%%%%%%%%%%%%%%%%%%%
\subsection[The models Tk]{\boldmath The models T$_k$}
\label{app:Tk}
Applying equation~\eqref{eq:recur:a} to $\sub{\Psi}{T_1}(\vec{r})=-GM\theta_0(\vec{r})$, we find for the Toomre-Miyamoto-Nagai models T$_k$
\begin{subequations}
	\label{eqs:pot:Tk}
\begin{align}
	\sub{\Psi}{T_1}(\vec{r})  &= -\theta_0,&
    \sub{\Phi}{T_1}(\vec{r})  &= -\frac1{\varrho};
    \\
	\sub{\Psi}{T_2}(\vec{r})  &= -\theta_0 - a\theta_1,&
    \sub{\Phi}{T_2}(\vec{r})  &= -\frac1{\varrho}\left[1 + \frac{a Z}{\varrho^2}\right];
    \\
	\sub{\Psi}{T_3}(\vec{r})  &= -\theta_0 - a\theta_1 - \frac{a^2}3\theta_2,&
    \sub{\Phi}{T_3}(\vec{r})  &= -\frac1{\varrho}\left[1 + \frac{a(Z-\tfrac13a)}{\varrho^2} + \frac{a^2 Z^2}{\varrho^4}\right];
    \\
	\sub{\Psi}{T_4}(\vec{r})  &= -\theta_0 - a\theta_1 - \frac{2a^2}5\theta_2 - \frac{a^3}{15}\theta_3,&
    \sub{\Phi}{T_4}(\vec{r})  &= -\frac1{\varrho}\left[1 + \frac{a(Z-\tfrac25a)}{\varrho^2} + \frac{\frac35a^2Z(2Z-a)}{\varrho^4} + \frac{a^3 Z^3}{\varrho^6}\right].
\end{align}
\end{subequations}
Here, we have set $GM=1$ and omitted the arguments of $\theta_n(\vec{r})$ for brevity. The input densities $\mathcal{R}_{\mathrm{T}_k}$ can be obtained from equation~\eqref{eq:density:MN}, giving (with $\zeta=z-a$)
\begin{align}
	\label{eq:Rho:Tk}
	\mathcal{R}_{\mathrm{T}_k}(\vec{r}) = -\p_z\Psi_{\mathrm{T}_k}(\vec{r}) + \zeta \p_z^2\Psi_{\mathrm{T}_k}(\vec{r}),
\end{align}
which implies that the $\mathcal{R}_{\mathrm{T}_k}$ follow the same recursion~\eqref{eq:recur:a} as the $\Psi_{\mathrm{T}_k}$. Using these relations, we find
\begin{subequations}
	\label{eqs:rho:Tk}
\begin{align}
	\sub{\mathcal{R}}{T_1}(\vec{r})&= \theta_1 + \zeta \theta_2,&
    \sub{\rho}{T_1}(\vec{r}) &= \frac{Mb^2}{4\upi} \left[\frac{a}{\varrho^3\zeta^3}+\frac{3 Z^2}{\varrho^5\zeta^2}\right];
    \\
	\sub{\mathcal{R}}{T_2}(\vec{r})&= \theta_1 + z\theta_2 +a \zeta \theta_3,&
    \sub{\rho}{T_2}(\vec{r}) &= \frac{3Mb^2}{4\upi}
		\left[\frac{\zeta^3+a^3}{\varrho^5\zeta^3} + \frac{5aZ^3}{\varrho^7\zeta^2} \right];
    \\
	\sub{\mathcal{R}}{T_3}(\vec{r})&= \theta_1 + z\theta_2 +\frac{a}3(3\zeta+a) \theta_3 + \frac{a^2}3\zeta\theta_4,&
    \sub{\rho}{T_3}(\vec{r}) &= \frac{Mb^2}{4\upi} \left[\frac{3}{\varrho^5} + \frac{5aZ^2(3\zeta^2-2a\zeta+a^2)}{\varrho^7\zeta^3} + \frac{35a^2Z^4}{\varrho^9\zeta^2} \right];
    \\
	\sub{\mathcal{R}}{T_4}(\vec{r})&= \theta_1 + z\theta_2 + \frac{a}5(5\zeta+2a)\theta_3 + \frac{a^2}{15}(6\zeta+a)\theta_4 + \frac{a^3}{15}\zeta\theta_5,&
    \sub{\rho}{T_4}(\vec{r}) &= \frac{Mb^2}{4\upi} \left[\frac{3}{\varrho^5} + \frac{15aZ}{\varrho^7} + \frac{7a^2Z^3(6\zeta^2-3a\zeta+a^2)}{\varrho^9\zeta^3} + \frac{63a^3Z^5}{\varrho^{11}\zeta^2} \right].
\end{align}
\end{subequations}
\citetalias{MiyamotoNagai1975} gave equivalent expressions for potential $\Phi$ and density $\rho$ of the models T$_{\text{1-3}}$.

%%%%%%%%%%%%%%%%%%%%%%%%%%%%%%%%%%%%%%%%%%%%%%%%%%%%%%%
\subsection[The models Vk]{\boldmath The models V$_{\!k}$}
\label{app:Vk}
These models are obtained via differentiation with respect to parameter $b$. More specifically, for the density of the models V$_{\!k}$ 
\begin{subequations}
\begin{align}
	\label{eq:rho:recur:l}
	\rho_{\mathrm{V}_{\!k}}(\vec{r};a,b) &= \hat{D}_{b,2} \cdot \rho_{\mathrm{T}_k}(\vec{r};a,b)
\end{align}
and hence by analogy
\begin{align}
	\Phi_{\mathrm{V}_{\!k}}(\vec{r};a,b) &= \hat{D}_{b,2} \cdot \Phi_{\mathrm{T}_k}(\vec{r};a,b),
\end{align}
which translate to, respectively,
\begin{align}
	\label{eq:recur:b}
	\mathcal{R}_{\mathrm{V}_{\!k}}(\vec{r}) = \left(3-\zeta\pdiff{}{z}\right)\mathcal{R}_{\mathrm{T}_k}(\vec{r})
	\quad\text{and}\quad
	\Psi_{\mathrm{V}_{\!k}}(\vec{r};a,b) = \left(1-\frac{b^2}{2\zeta}\pdiff{}{z}\right) \cdot \Psi_{\mathrm{T}_k}(\vec{r};a)
\end{align}
\end{subequations}
for the input models, which also between them follow equation~\eqref{eq:recur:a} but not~\eqref{eq:Rho:Tk}. We find
\begin{subequations}
	\label{eqs:pot:Vk}
\begin{align}
	\sub{\Psi}{V_{\!1}}(\vec{r})   &= \sub{\Psi}{T_1} - \frac{b^2}{2\zeta}\theta_1,&
    \sub{\Phi}{V_{\!1}}(\vec{r})   &= -\frac1{\varrho}\left[1 + \frac{b^2Z}{2\varrho^2\zeta}\right];
    \\
	\sub{\Psi}{V_{\!2}}(\vec{r})   &= \sub{\Psi}{T_2} - \frac{b^2}{2\zeta}\left(\theta_1+a\theta_2\right),&
	\sub{\Phi}{V_{\!2}}(\vec{r})   &= -\frac1{\varrho}\left[1 + \frac{aZ+\tfrac12b^2}{\varrho^2} + \frac{3ab^2Z^2}{2\varrho^4\zeta} \right];
    \\
	\sub{\Psi}{V_{\!3}}(\vec{r})   &= \sub{\Psi}{T_3} - \frac{b^2}{2\zeta}\left(\theta_1+a\theta_2 + \frac{a^2}3\theta_3\right),&
	\sub{\Phi}{V_{\!3}}(\vec{r})   &= -\frac1{\varrho}\left[1+\frac{a(Z-\frac13a)+\frac12b^2}{\varrho^2} + \frac{aZ(aZ+\frac32b^2)}{\varrho^4} + \frac{\frac52a^2b^2Z^3}{\varrho^6\zeta}\right];
    \\
	\sub{\Psi}{V_{\!4}}(\vec{r})   &= \sub{\Psi}{T_4} - \frac{b^2}{2\zeta}\left(\theta_1 + a\theta_2 + \frac{2a^2}5\theta_3 + \frac{a^3}{15}\theta_4\right),&
	\sub{\Phi}{V_{\!4}}(\vec{r})   &= -\frac1{\varrho}\left[1 + \frac{a(Z-\tfrac25a)+\tfrac12b^2}{\varrho^2} + \frac{3a[\tfrac15aZ(2Z-a)+\tfrac12b^2(Z-\tfrac15a)]}{\varrho^4} + \right. \nonumber \\
    &&& \phantom{-\frac1{\varrho}+++}\left. + \frac{a^2 Z^2(aZ+3b^2)}{\varrho^6} + \frac{\tfrac72b^2a^3z^4}{\varrho^8\zeta}\right]
\end{align}
\end{subequations}
for the gravitational potentials and
\begin{subequations}
	\label{eqs:rho:Vk}
\begin{align}
	\sub{\mathcal{R}}{V_{\!1}}(\vec{r}) &= 3 \sub{\mathcal{R}}{T_1} + \zeta^2 \theta_3,&
	\sub{\rho}{V_{\!1}}(\vec{r}) &= \frac{3Mb^4}{8\upi\zeta^5}\left[\frac{a}{\varrho^3} + \frac{3aZ\zeta}{\varrho^5} + \frac{5Z^3\zeta^2}{\varrho^7} \right];
    \\
	\sub{\mathcal{R}}{V_{\!2}}(\vec{r}) &= 3 \sub{\mathcal{R}}{T_2} + \zeta^2 \big(\theta_3 + a \theta_4\big),&
	\sub{\rho}{V_{\!2}}(\vec{r}) &= \frac{3Mb^4}{8\upi\zeta^5}\left[\frac{3a^3}{\varrho^5} + \frac{5Z^2\zeta(\zeta^2-2a\zeta+3a^2)}{\varrho^7} + \frac{35aZ^4\zeta^2}{\varrho^9}\right];
    \\	
	\sub{\mathcal{R}}{V_{\!3}}(\vec{r}) &= 3 \sub{\mathcal{R}}{T_3} + \zeta^2 \left(\theta_3 + a \theta_4 + \frac{a^2}3  \theta_5\right),&
	\sub{\rho}{V_{\!3}}(\vec{r}) &= \frac{5Mb^4}{8\upi\zeta^5}\left[\frac{3(\zeta^5+a^5)}{\varrho^7} + \frac{7aZ^3\zeta(3\zeta^2-4a\zeta+3a^2)}{\varrho^9} + \frac{63a^2Z^5\zeta^2}{\varrho^{11}}\right];
    \\
	\sub{\mathcal{R}}{V_{\!4}}(\vec{r}) &= 3 \sub{\mathcal{R}}{T_4} + \zeta^2 \left(\theta_3 + a \theta_4 + \frac{2a^2}5  \theta_5 + \frac{a^3}{15}\theta_6\right),&
    \sub{\rho}{V_{\!4}}(\vec{r}) &= \frac{3Mb^4}{8\upi\zeta^5} \left[\frac{5\zeta^5}{r^7} + \frac{7az^2(5\zeta^4-4a\zeta^3+3a^2\zeta^2-2a^3\zeta+a^4)}{r^9} + \right. \nonumber \\
    &&& \phantom{\frac{3Mb^4}{8\upi\zeta^5}++}\left. + \frac{63a^2z^4\zeta(2\zeta^2-2a\zeta+a^2)}{r^{11}} + \frac{231a^3z^6\zeta^2}{r^{13}}\right].
\end{align}
\end{subequations}
for the densities.

%%%%%%%%%%%%%%%%%%%%%%%%%%%%%%%%%%%%%%%%%%%%%%%%%%%%%%%
\section{Vertical (face-on) projection}
\label{app:proj:z}
The vertical projection of the axisymmetric models T$_k$ and V$_{\!k}$ (we deal with the barred versions at the end of this Appendix)
\begin{align}
	\label{eq:sigma:int}
	\Sigma(R) = 2\int_0^\infty \rho(R,z) \diff z
\end{align}
is conveniently computed using the substitution
\begin{align}
	\label{eq:Sigma:subst}
	t = \frac{z}{\sqrt{z^2+q^2}},\quad
	z = \frac{qt}{\sqrt{1-t^2}},\quad
	\frac{\diff z}{\diff t} = \frac{(z^2+q^2)^{3/2}}{q^2} = \frac{q}{(1-t^2)^{3/2}},
\end{align}
where $q$ is a parameter. When using $q=b$, then $(b^2/\zeta^3)\diff z = \diff t$, which accounts for the factor $\zeta^{-3}$ in the density $\rho$ of all the models. Unfortunately, this promising approach fails at $R\gg a$ for all models (except V$_{\!1}$), since they contain spheroidal components, which lack the factor $\zeta^{-3}$ and at $R\gg a$ dominate the integral for $\Sigma$, i.e.\ the density decays less steeply with $z$ than anticipated by the substitution.

Therefore, we now develop an alternative for computing the vertical projection. For models T$_k$, the input model satisfies $\vec{\nabla}^2\Psi(\vec{r})=0$ and the density can be written as
\begin{align}
	4\upi G\rho(\vec{r}) = \pdiff{}{z} \left(\frac{z}{\zeta} \pdiff{\Psi(\vec{\varrho})}{Z}\right) - \frac{\p^2\Psi(\vec{\varrho})}{\p Z^2},
\end{align}
equivalent to equation~\eqref{eq:density:MN}. When inserting this into equation~\eqref{eq:sigma:int}, the first term integrates to zero, such that
\begin{align}
	\Sigma(R) = -\frac1{2\upi G} \int_0^\infty \frac{\p^2\Psi(\vec{\varrho})}{\p Z^2}\diff z.
\end{align}
For model T$_1$, $\Psi=-GM\theta_0(\vec{\varrho})$ and $\p^2\Psi/\p Z^2=-GM\theta_2(\vec{\varrho})$. In general, for any of the axisymmetric disc models we can write
\begin{align}
	\label{eq:sigma:int:sig}
	\Sigma(R) = \frac{M}{2\upi} \int_0^\infty \sigma(\vec{\varrho}) \diff z.
\end{align}
which can be evaluated using the substitution~\eqref{eq:Sigma:subst}. The functions $\sigma$ can be obtained for all the models via differentiation w.r.t.\ parameters $a$ or $b$ and follow the same recursions~\eqref{eq:recur:a} and~\eqref{eq:recur:b} as the input potentials $\Psi$. We find (omitting the argument of $\sigma(\vec{r})$ for brevity)
\begin{subequations}
\begin{align}
	\sub{\sigma}{T_1} &= \theta_2 &
				&= - \frac1{r^3} + \frac{3z^2}{r^5},\\
	\sub{\sigma}{T_2} &= \theta_2 + a \theta_3 &
				&= - \frac1{r^3} + \frac{3z(z-3a)}{r^5} + \frac{15az^3}{r^7},\\
	\sub{\sigma}{T_3} &= \theta_2 + a \theta_3 + \frac{a^2}3 \theta_4 &
				&= - \frac1{r^3} + \frac{3(z^2-3az+a^2)}{r^5} + \frac{15az^2(z-2a)}{r^7} + \frac{35a^2z^4}{r^9},\\
	\sub{\sigma}{T_4} &= \theta_2 + a \theta_3 + \frac{2a^2}5 \theta_4 + \frac{a^3}{15}\theta_5 &
				&= - \frac1{r^3} + \frac{3(z^2-3az+\tfrac65a^2)}{r^5} + \frac{3az(5z^2-12az+5a^2)}{r^7} + \frac{14a^2z^3(3z-5a)}{r^9} + \frac{63a^3z^5}{r^{11}},\\
	\sub{\sigma}{V_{\!1}} - \sub{\sigma}{T_1} &= \frac{b^2}{2\zeta} \theta_3 &
				&= \frac{b^2}{2} \left[- \frac{9z}{r^5\zeta} +\frac{15z^3}{r^7\zeta} \right],\\
	\sub{\sigma}{V_{\!2}} - \sub{\sigma}{T_2} &= \frac{b^2}{2\zeta} \left[\theta_3 + a\theta_4 \right] &
				&= \frac{b^2}{2} \left[- \frac{9}{r^5} + \frac{15z^2(z-6a)}{r^7\zeta} + \frac{105az^4}{r^9\zeta}\right],\\
	\sub{\sigma}{V_{\!3}} - \sub{\sigma}{T_3} &= \frac{b^2}{2\zeta} \left[\theta_3 + a\theta_4 +\frac{a^2}3\theta_5 \right] &
				&= \frac{b^2}{2} \left[ -\frac{9}{r^5} + \frac{15z(z-5a)}{r^7} + \frac{35az^3(3z-10a)}{r^9\zeta} + \frac{315a^2z^5}{r^{11}\zeta} \right]\\
	\sub{\sigma}{V_{\!4}} - \sub{\sigma}{T_4} &= \frac{b^2}{2\zeta} \left[\theta_3 + a\theta_4 +\frac{2a^2}5\theta_5 + \frac{a^3}{15}\theta_6 \right] & 
				&= \frac{b^2}{2} \left[ -\frac{9}{r^5} + \frac{15(z^2-5az+a^2)}{r^7} + \frac{105az^2(z-3a)}{r^9} + \frac{189a^2z^4(2z-5a)}{r^{11}\zeta} + \frac{693a^3z^6}{r^{13}\zeta}\right].
\end{align}
\end{subequations}
In practice, we compute the integrals~\eqref{eq:sigma:int:sig} via Gau\ss{}-Legendre integration with 64 points, using the substitution~\eqref{eq:Sigma:subst} with $q^2=x^2+y^2+s^2$. This is typically sufficient for a relative error of $10^{-8}$ or better. The convolution with the needle density to obtain the barred models is accomplished by the replacements~\eqref{eq:bar:Phi,rho} in the relations for $\sigma(\vec{\varrho})$.

%%%%%%%%%%%%%%%%%%%%%%%%%%%%%%%%%%%%%%%%%%%%%%%%%%%%%%%
\label{lastpage}
\end{document}